\documentclass[%
  a4paper,
  amsfonts,amssymb,amsmath,
  reprint,
  superscriptaddress,        
  showkeys,nofootinbib,twoside
]{revtex4-2}
\usepackage[english]{babel}
\usepackage[utf8]{inputenc}
\usepackage{amsthm}
\usepackage{siunitx}
\usepackage{comment}
\usepackage{graphicx}
\usepackage{subcaption}
\usepackage[colorinlistoftodos, color=green!40, prependcaption]{todonotes}
\usepackage{lineno}
\usepackage{amsthm}
\usepackage{mathtools}
\usepackage{physics}
\usepackage{xcolor}
\usepackage{graphicx}
\usepackage[left=23mm,right=13mm,top=35mm,columnsep=15pt]{geometry} 
\usepackage{adjustbox}
\usepackage{placeins}
\usepackage[T1]{fontenc}
\usepackage{lipsum}
\usepackage{csquotes}
\usepackage[pdftex, pdftitle={Article}, pdfauthor={Author}]{hyperref} 
\begin{document}
\title{Electrical Control of Optically Active Single Spin Qubits in ZnSe}

\author{Amirehsan Alizadehherfati}
\email{herfati@umd.edu}
\affiliation{Institute for Research in Electronics and Applied Physics and Joint Quantum Institute, University of Maryland, College Park, Maryland 20742, USA}
\affiliation{Department of Electrical and Computer Engineering, University of Maryland, College Park, MD 20742, USA}

\author{Yuxi Jiang}
\affiliation{Institute for Research in Electronics and Applied Physics and Joint Quantum Institute, University of Maryland, College Park, Maryland 20742, USA}
\affiliation{Department of Electrical and Computer Engineering, University of Maryland, College Park, MD 20742, USA}

\author{Nils von den Driesch}
\affiliation{Peter-Gr\"unberg Institute (PGI-9 \& PGI-10), Forschungszentrum J\"ulich GmbH, 52428 J\"ulich, Germany}
\affiliation{JARA-Fundamentals of Future Information Technology, Forschungszentrum J\"ulich and RWTH Aachen University, 52062 Aachen, Germany}

\author{Christine Falter}
\affiliation{Peter-Gr\"unberg Institute (PGI-9 \& PGI-10), Forschungszentrum J\"ulich GmbH, 52428 J\"ulich, Germany}
\affiliation{JARA-Fundamentals of Future Information Technology, Forschungszentrum J\"ulich and RWTH Aachen University, 52062 Aachen, Germany}

\author{Yurii Kutovyi}
\affiliation{Peter-Gr\"unberg Institute (PGI-9 \& PGI-10), Forschungszentrum J\"ulich GmbH, 52428 J\"ulich, Germany}
\affiliation{JARA-Fundamentals of Future Information Technology, Forschungszentrum J\"ulich and RWTH Aachen University, 52062 Aachen, Germany}

\author{Jasvith Raj Basani}
\affiliation{Institute for Research in Electronics and Applied Physics and Joint Quantum Institute, University of Maryland, College Park, Maryland 20742, USA}
\affiliation{Department of Electrical and Computer Engineering, University of Maryland, College Park, MD 20742, USA}

\author{Amirehsan Boreiri}
\affiliation{Institute for Research in Electronics and Applied Physics and Joint Quantum Institute, University of Maryland, College Park, Maryland 20742, USA}
\affiliation{Department of Electrical and Computer Engineering, University of Maryland, College Park, MD 20742, USA}

\author{Alexander Pawlis}
\affiliation{Peter-Gr\"unberg Institute (PGI-9 \& PGI-10), Forschungszentrum J\"ulich GmbH, 52428 J\"ulich, Germany}
\affiliation{JARA-Fundamentals of Future Information Technology, Forschungszentrum J\"ulich and RWTH Aachen University, 52062 Aachen, Germany}

\author{Edo Waks}
\email{edowaks@umd.edu}
\affiliation{Institute for Research in Electronics and Applied Physics and Joint Quantum Institute, University of Maryland, College Park, Maryland 20742, USA}
\affiliation{Department of Electrical and Computer Engineering, University of Maryland, College Park, MD 20742, USA}


\begin{abstract}
Electrons bound to shallow donors in ZnSe quantum wells are promising candidates for optically addressable spin qubits and single-photon sources. However, their optical coherence and indistinguishability are often limited by spectral broadening arising from charge fluctuations in the local environment. Here, we report electrical control of single donor qubits in ZnSe quantum wells. The applied field induces a DC Stark shift that tunes the emission energy over a range exceeding 30 times the inhomogeneous linewidth, effectively compensating for emitter-to-emitter variations. Concurrently, the field stabilizes trap occupancy, yielding a twofold reduction in optical linewidth and the suppression of spectral wandering. A statistical model based on trap dynamics qualitatively reproduces these observations and elucidates the mechanism of field-assisted charge noise suppression. Our results identify electrical control as a versatile pathway to significantly improve optical and spin addressability.

\end{abstract}


\maketitle

\section{Introduction}

Optically addressable qubits in semiconductors are attractive building blocks for quantum network and quantum memory applications \cite{doi:10.1126/science.abg1919,Knaut2024Entanglement,awschalom2025challenges}. Among them, the coupled electron-nuclear spin system associated with the donors offers a multilevel platform with potentially long coherence times \cite{Linpeng2018Coherence,Karin2016Longitudinal,Viitaniemi2022Coherent}.
Direct bandgap host crystals such as ZnSe provide efficient radiative transitions \cite{Qiao2024,Kutovyi2022Efficient,Kim2012Semiconductor,doi:10.1021/acsphotonics.3c01540} and, through isotopic purification \cite{Pawlis2019MBE,Kopteva2019-yp,Kirstein2021Extended}, an almost nuclear spin free environment for the donors. In particular, shallow donors in ZnSe quantum wells have emerged as a promising platform \cite{De-Greve2010-fi,Karasahin2022,doi:10.1021/nl303663n}. However, their potential is limited due to the environmental decoherence \cite{Kuhlmann2013ChargeSpinNoise,Kim2014-od}. Static disorder within the host material gives rise to inhomogeneous broadening, while dynamic charge noise, originating from charge carriers trapped at nearby defects and interfaces, induces spectral diffusion via the Stark effect . These processes introduce dephasing channels that degrade photon indistinguishability, thereby limiting Hong-Ou-Mandel visibility \cite{Sanaka2012Entangling,Sanaka2009Indistinguishable} and reducing spin coherence times.

As a solution, electrical control provides a versatile approach to tune the optical properties of semiconductor quantum emitters \cite{doi:10.1126/science.278.5346.2114}. In particular, applied electric fields can mitigate inhomogeneous broadening caused by variations in local strain and charge environment by tuning disparate emitters into resonance via the DC Stark effect \cite{Patel2010RemoteHOM,PhysRevLett.108.043604,bushmakin2025twophotoninterferencephotonsremote,PhysRevLett.131.033606,zhai2022quantum}. Furthermore, external electric fields can stabilize the fluctuations in local electric field, reducing spectral wandering caused by charge noise \cite{Anderson2019Electrical,Somaschi2016NearOptimal}. These fields are commonly applied using device structures such as p-i-n diodes \cite{Kuhlmann2015TransformLimited,PhysRevLett.104.047402,zeledon2025minute}, Schottky diodes \cite{steidl2025single}, or surface electrodes \cite{Aghaeimeibodi2021Electrical,DeSantis2021Stark}. Despite these advantages, the difficulty of p-doping ZnSe has prevented the realization of electric-field control for single impurities \cite{PhysRevLett.74.1131,PhysRevB.45.10965,doi:10.1021/acsaelm.4c01104}.

In this Letter, we demonstrate electrical control of single donor-bound excitons in ZnSe quantum wells using laterally applied fields from patterned surface electrodes. This configuration enables local tuning of the emission energy by polarizing the bound exciton. In addition to spectral tuning, we find the lateral field actively stabilizes the local charge environment, manifesting as a two fold narrowing of the emission linewidth. We attribute this narrowing to the suppression of spectral wandering arising from trap-induced charge fluctuations, a mechanism corroborated by similar behavior under a weak, above-band optical pump. A statistical model based on trap charge dynamics qualitatively reproduces the observed line broadening and subsequent narrowing under both optical and electrical stabilization. These results establish electric fields as a powerful tool for simultaneously achieving precise spectral tuning and stabilizing the charge environment of quantum emitters.

\section{Initial Characterization}

\begin{figure}[t]
  \centering
  \includegraphics[width=\columnwidth]{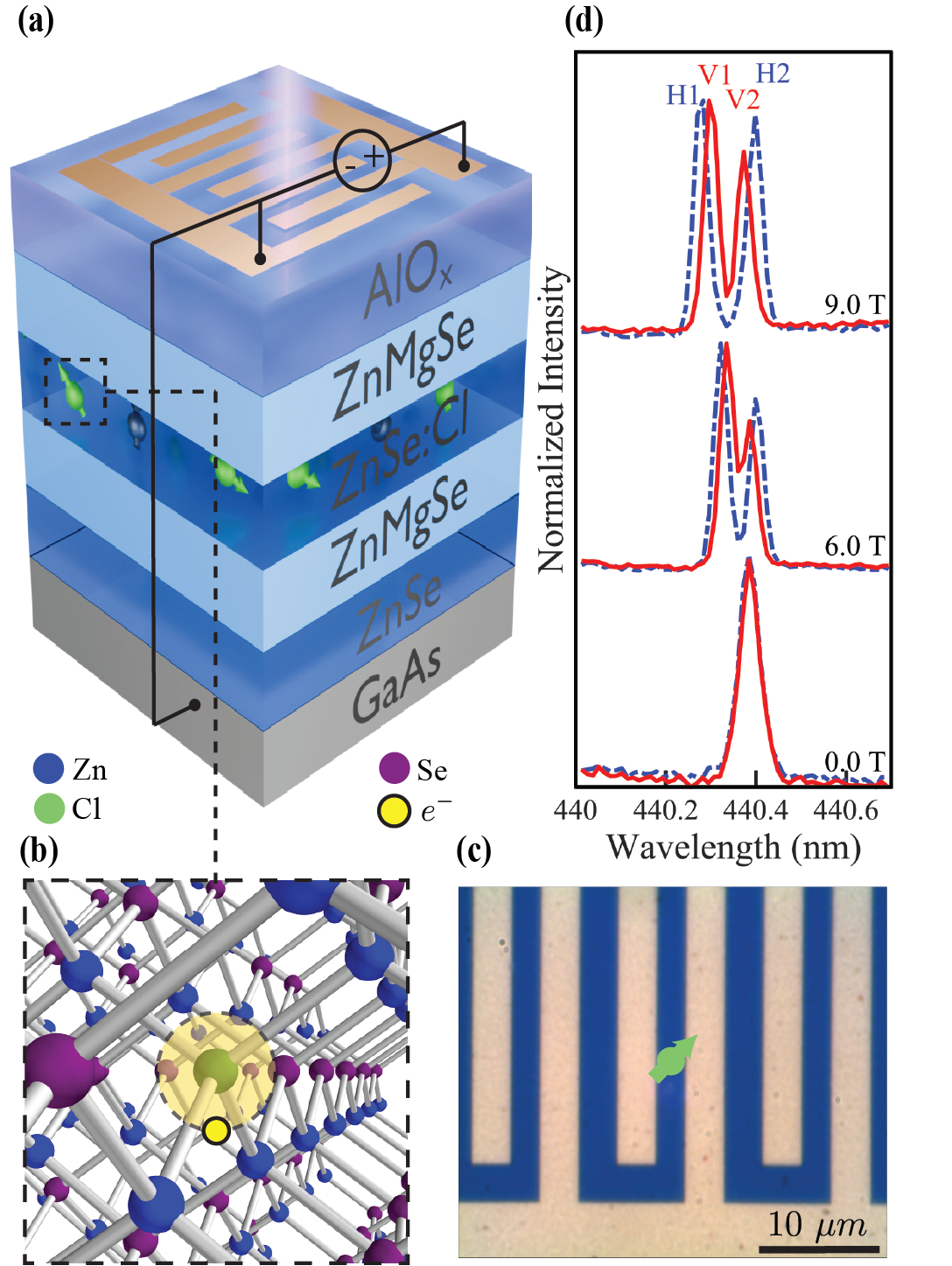}
  \caption{\textbf{Device illustration and magneto spectroscopy characterization.} 
  (a) Schematic of device geometry, consisting of ZnMgSe/ZnSe:Cl/ZnMgSe quantum well, ZnSe buffer, AlO$_x$ capping and GaAs substrate, with electrodes on the top and connections. (b) Crystal structure Zn(blue)Se(pink) with Cl(green) as a substitutional donor and bound electron (yellow). (c) Optical microscope image of the fabricated electrodes. The gold-colored regions highlight the patterned electrodes, designed for lateral electric field application. (d) Photoluminescence spectrum of donor-bound exciton in Voigt configuration shows four distinct peaks with two orthogonal polarization.
}
\label{fig:Fig0}
\end{figure}

Fig. \ref{fig:Fig0}(a) provides an illustration of our device. The initial sample consists of a GaAs‐substrate–based ZnMgSe/ZnSe/ZnMgSe quantum well incorporating a Cl $\mathrm{\delta}$-doped layer at its center \cite{Ohkawa1987}. Based on first-principles calculations and growth conditions, we expect the majority of chlorine atoms to occupy substitutional Se sites, acting as shallow donors $Cl_{Se}^{+1}$ in the ZnSe lattice, as illustrated in Fig. \ref{fig:Fig0}(b) \cite{dosSantos2011,Poykko1998,Wu2022,Wu2022NativeZnSe,10.1063/5.0265591}. The donor-bound electron acts as a spin $1/2$ qubit and is optically accessible through radiative recombination of bound exciton \cite{Dean1981-sh,Steiner1985-ti,Merz1972OpticalProperties}. To apply a lateral electric field across individual impurity sites distributed inside well, we pattern interdigitated electrodes via a standard metal liftoff process (see the Methods section A) on the surface. A potential difference is applied at two opposing electrodes. The top $\mathrm{AlO}_{\mathrm{x}}$ layer prevents current flow through the quantum well, suppressing leakage current and Joule heating. Fig. \ref{fig:Fig0}(c) shows an optical microscope image of the device from the top.

All experiments are performed at \SI{3.6}{\K} in a closed-loop cryostat using a home-built confocal microscope. The optical setup includes separate ports for above-band excitation, resonant excitation, and signal collection, with polarization optics arranged in a cross-polarization scheme to suppress resonant laser reflection. A long-pass filter (cutoff $\approx\ $\SI{438}{\nano\metre}) is used during long-exposure resonant measurements. Spectroscopy is conducted with a grating spectrometer offering \SI{0.02}{\nano\metre} (\SI{128}{\micro\eV}) resolution. More details about the setup can be found in the Methods section B.

We initially characterize the device using photoluminescence spectroscopy under above-band excitation at \SI{405}{\nano\metre}. We scan the regions between the electrodes to locate photoluminescent impurity sites. At various sites, we observe isolated sharp emission peaks corresponding to $D^0X$ to $D^0$ transition of single Cl donors. Previous studies on the same quantum emitter have shown antibunching from this transition \cite{Karasahin2022,jiang2024generation,Kutovyi2022Efficient}. 

To confirm the origin of the emission, we perform polarization-resolved magnetospectroscopy. Fig. \ref{fig:Fig0}(d) shows the photoluminescence signal from the emitter for magnetic fields up to \SI{9}{\T} perpendicular to the sample growth direction (Voigt configuration). The external field lifts the degeneracy of the electron and hole spin states in the bound exciton complex and produces four allowed optical transitions \cite{astakhov2002binding}. Each pair of transitions has a specific linear polarization, and together they form two orthogonal polarization bases, shown as H and V \cite{Greilich2012-ic,De-Greve2010-fi}. From the Zeeman-split lines we extract effective g-factors corresponding to each polarization (see Supplemental Material section I). Based on the anticipated level structure, we find the ground state electron g-factor of $g_e = 1.19\pm 0.02$, and excited state heavy hole g-factor of $g_h = 0.09\pm 0.02$. Both values are in close agreement with previous studies on single Fluorine donors \cite{Greilich2012-ic,Kopteva2019-yp,De-Greve2010-fi}. For all subsequent measurements, we focus on the electrical properties of this transition at zero magnetic field.

\section{Spectral Tuning}
\begin{figure}[t]
  \centering
  \includegraphics[width=\columnwidth]{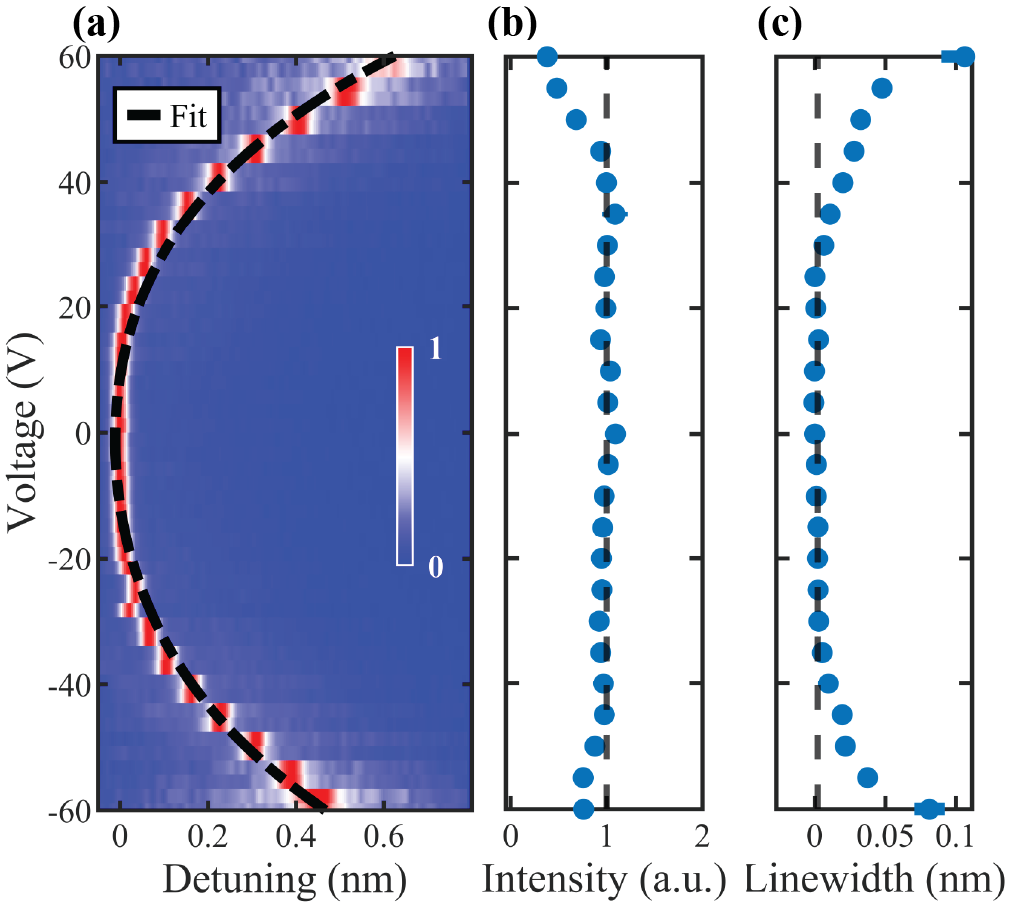}
  \caption{\textbf{Characterization of the Stark shift in a single bound exciton.} (a) Photoluminescence bias map. (b) Integrated intensity as a function of the applied bias. The dashed line indicates the average intensity of emission in the absence of electric field. (c) Full width at half maximum of the emission line as a function of the applied bias. The dashed line shows the average linewidth of emission in the absence of bias voltage.
}
\label{fig:Fig1}
\end{figure}

Fig. \ref{fig:Fig1}(a) shows the response of the bound exciton line as a function of applied bias voltage. The emission undergoes a voltage-dependent spectral shift towards longer wavelengths. We attribute this shift to the quantum-confined Stark effect \cite{fu1987electric}, in which the applied electric field polarizes the bound exciton by spatially separating the electron and hole \cite{brum1985electric}, inducing a dipole moment and reducing the recombination energy, resulting in a redshift of the emission \cite{miller1985electric,PhysRevLett.53.2173}. The dashed black line is a numerical fit to the central wavelength of the emission signal, containing polynomial terms up to fourth degree. We performed this fit by initially fitting a Voigt profile to each spectrum, extracting emission properties including central wavelength, total intensity and linewidth. From this fit, we obtained a polarizability of $\beta = (2.6 \pm 0.2)\times10^{-6}\ \mathrm{meV}\,(\mathrm{cm}/\mathrm{kV})^{2}$ and a permanent dipole moment of $d = (1.0 \pm 0.8)\times10^{-4}\ \mathrm{meV}\,(\mathrm{cm}/\mathrm{kV})$. The first order contribution is relatively small, showing the lack of a notable permanent dipole moment. Additionally, we find $(4.1 \pm 1.1)\times10^{-10}\ \mathrm{meV}\,(\mathrm{cm}/\mathrm{kV})^{3}$ and $(1.1 \pm 0.2)\times10^{-12}\ \mathrm{meV}\,(\mathrm{cm}/\mathrm{kV})^{4}$  for higher-order dependency of Stark shift as a function of the electric field. From this analysis, we can deduce that the quadratic and quartic terms are dominant, which we attribute to the centrosymmetric nature of excitonic transitions in a quantum well \cite{mendez1989optical}. Details of our finite-element simulations used to calculate the internal field from applied voltage bias are provided in the Supplemental Material section IV.A.

\begin{figure*}[t]
  \includegraphics[width=\textwidth]{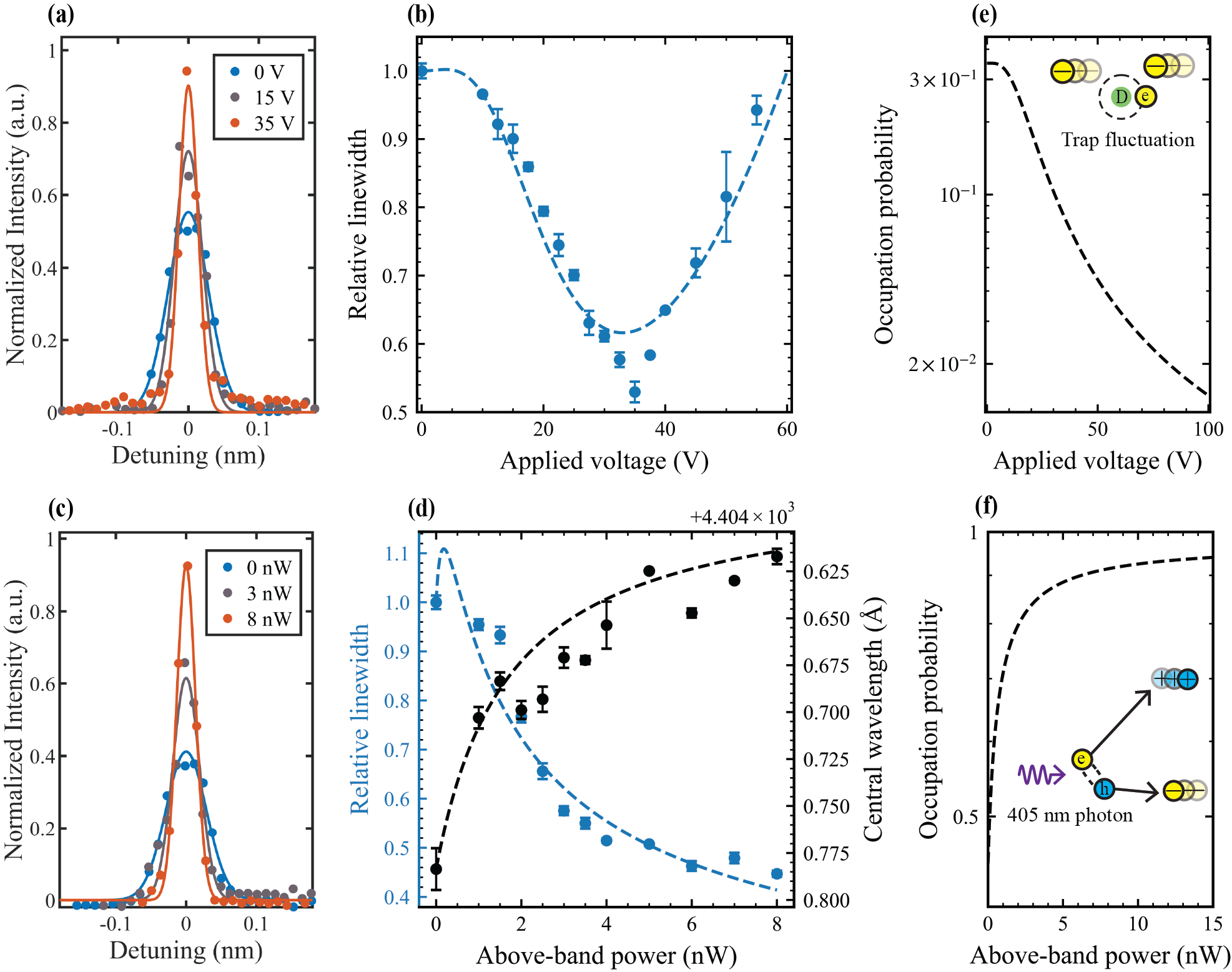}
  \caption{\textbf{Manipulation of charge environment around donor site.} (a)  Emission spectrum under resonant excitation of the free exciton line for different electric fields. (b) The narrowing effect from the electric field suggests an optimal field of around 35 V is suitable to stabilize the bound exciton emission. The theoretical curve corresponds to about 18  traps distributed in the region of \SI{3}{\nano m} to \SI{5}{\nano m} from donor with average trap depth of \SI{0.3}{eV} and initial occupancy of 0.35. We extract these values based on a numerical fit constrained by the donor's effective Bohr radius and typical defect densities in ZnSe.  (c)  Emission spectra recorded for varying weak above-band excitation powers in the absence of an electric field. (d) Dynamics of linewidth and central wavelength as a function of different above-band intensities. The theoretical curves correspond to an average saturation power of 1 nW and an initial trap occupancy of about 0.4. We extract these values based on a numerical fit to a statistical model.  (e) The electric field stabilizes the fluctuating traps by polarizing the occupancy probability toward 0.  (f) The process of filling trap states and eliminating the fluctuating charge environment through free carrier injection pushes occupancy toward unity.  
}
\label{fig:Fig3}
\end{figure*}

We next examine how the electric field affects the brightness of the impurity-bound exciton. Fig. \ref{fig:Fig1}(b) plots the total intensity of emission as a function of voltage bias. For the absolute biases lower than 40 V, the intensity level is stable, indicating that the electric field has a negligible destructive effect on the emission. At higher fields, we observe a decrease in the intensity, indicating a quenching behavior which determines the fundamental tuning range of our device. We attribute the quenching to tunneling of the charge from the impurity site and dissociation of the bound exciton in high fields \cite{PhysRevLett.55.2610,wu1988exciton}. Supplemental Material section IV.B discusses the dynamics of free-exciton and free-trion emission under an electric field, showing that the bound exciton is significantly more stable than the negatively charged trion due to the strong confinement of the ground-state electron at the impurity site \cite{PhysRevB.55.R1970}.

In the next step, we investigate the behavior of the linewidth as a function of the applied electric field. Fig.~\ref{fig:Fig1}(c) shows the linewidth of bound exciton emission as a function of voltage. We extract the linewidth by calculating the full width at half maximum of the Voigt fit. For voltages below 35 V, the linewidth remains constant and is limited by the spectrometer resolution. This range, in good agreement with Fig.~\ref{fig:Fig1}b, is the regime in which the device operates at optimal capacity. Once the applied voltage exceeds 35 V, we observe linewidth start to broaden, which we attribute to increased sensitivity of the emitter to background charge fluctuations \cite{DeSantis2021Stark} and possible heating from the electrodes. 

While high fields eventually induce broadening, we observe an intermediate regime where the field significantly reduces the bound exciton linewidth by actively suppressing spectral diffusion. We investigate this stabilization mechanism in the following section

\section{Manipulation of Charge Environment}
\label{sec:develop}

To isolate the effect of the electric field on the linewidth, we need to change the excitation method. Linewidth is strongly determined by local charge environment, and above-band pumping generates a significant background of free carriers, perturbing the local charge landscape. Therefore, photoluminescence spectroscopy cannot show any subtle effects of electric-field-induced charge stabilization around the emitter. Although directly driving the bound-exciton resonance is possible through the observed polarization selectivity (see Supplemental Material section II), cross-polarization is limited due to unwanted reflection from metal surfaces.

As an alternative approach, we perform photoluminescence-excitation spectroscopy by resonantly pumping the free exciton and measuring the bound exciton emission. A narrow-linewidth laser ($\approx 90 \mathrm{kHz}$) is tuned to selectively pump the free exciton on resonance, which induces the bound exciton emission through a cascaded process. Before each measurement, we compensate for the Stark shift of the free exciton by tuning our laser back on the free-exciton resonance (see Supplemental Material section III.B)

Fig. \ref{fig:Fig3}(a) shows the photoluminescence-excitation spectra for different values of voltage, all centered at the same wavelength.  In the absence of an electric field, the signal initially exhibits about three times broader linewidth than the spectrometer limited resolution photoluminescence signal. In the presence of the field, the linewidth undergoes a clear reduction as the electric field increases. We repeat this measurement over a broader range of applied voltages while keeping the resonant power fixed (\SI{50}{\nano W}). Fig. \ref{fig:Fig3}(b) shows the extracted linewidth as a function of applied biases. We extract the linewidth values from a Voigt model fit to the spectrum at each bias and then normalize them to the zero-field intensity value. The trend shows an initial narrowing dynamic followed by a monotonic broadening effect. This behavior reveals an optimal field around \SI{35}{V}, at which the emission linewidth reaches a minimum at $50\%$ of the initial linewidth. Beyond this point, the observed broadening is consistent with previous photoluminescence observations, showing sensitivity to background charge fluctuations.  The initial broadening and its subsequent suppression with applied field can be attributed to trap states \cite{karczewski1994deep,hernandez1996investigation} and suppression of the resulting charge fluctuations in the emitter’s local environment \cite{Anderson2019Electrical,Somaschi2016NearOptimal,houel2012probing}. The dashed curve shows the best fit based on a statistical model, elaborated upon later in this section.

As further evidence on the suppression of trap states, we perform photoluminescence excitation spectroscopy while adding a weak above-band pump in the absence of an electric field. The power level used in this experiment is approximately two orders of magnitude lower than the saturation level (see Supplemental Material section V), resulting in negligible direct emission from above-band. Fig. \ref{fig:Fig3}(c) shows the photoluminescence-excitation signal for different values of the above-band power, all centered at the same wavelength. The linewidth exhibits clear narrowing as the above-band power increases.  We repeat the measurement across a broader span of above-band powers while keeping the resonant power fixed. Fig.~\ref{fig:Fig3}(d) shows the extracted central wavelength (black dots) and normalized linewidth (blue dots) as a function of weak above-band power. The emitter is initially broad and undergoes a clear reduction in linewidth until it reaches the spectrometer's limited resolution, almost matching the photoluminescence signal and reaching a final value around $40\%$ of initial linewidth. Concurrently, the central wavelength shifts toward the blue in response to increasing power, reaching a saturation plateau. We extract the central wavelength and linewidth through a Voigt fit to the spectra at each above-band power and normalize linewidths to the zero-above-band power value.  We attribute both observations to the generation of excess carriers under above-band illuminations that fill nearby trap states. This effect results in suppression of the built-in DC Stark effect and electric-field fluctuations, manifesting as reduced spectral diffusion and linewidth narrowing \cite{Majumdar2011-yp,Gazzano2018-ki,Chen2016-pj,PhysRevB.93.195316}. The dashed curves show the extracted best fit for relative linewidth and central wavelength based on a statistical model described in the subsequent discussion. We note that in previous studies \cite{Karasahin2022,jiang2024generation}, a weak above-band pump was necessary in order to observe any photoluminescence excitation signal from the donor-bound exciton. The donors were initially ionized, and the above-band light was required to supply the electron ground state. Observing the direct photoluminescence excitation signal here, without such pumping, suggests that the surface passivation and oxide capping are effective in partially stabilizing the donors.

Based on these results, we develop a statistical model to describe the possible physics behind the effect of fluctuating traps on the emitter's optical properties.  Each trap is considered as a two-state Markov process (charged or empty) with approximately identical capture $k^+$ and release rates $k^-$ among traps. When charged, the trap creates a local electric field at the emitter site that Stark-shifts the transition. Random switching across many traps produces a quasi-static distribution of detunings that is well approximated by a Gaussian profile \cite{kubo1969stochastic,w1954mathematical}. Convolution with the Fourier-limited Lorentzian of the transition yields the observed Voigt line \cite{armstrong1967spectrum}. To reflect device confinement, we assume an isotropic distribution of traps in the two-dimensional quantum-well plane and derive analytical expressions for the linewidth $2\sqrt{2\ln{2}}\ \sigma_{\delta\omega}$ and central wavelength $\mu_{\delta\omega}$ of the Gaussian profile in terms of the capture and release rates. 

\begin{widetext}
\begin{align}
\mu_{\delta\omega} =& \beta\Big(E_0^2+p\,S_2\Big)\\
\sigma_{\delta\omega}^2
=&\beta^2\left[
p(1-p)\big(S_4+2E_0^2 S_2\big)\;+\;p^2(1-p^2)\big(S_2^2-S_4\big)
\right]
\label{eq:sigma_bias_final}
\end{align}
\end{widetext}

where $p = k^+/(k^- + k^+)$ is the probability of trap occupation, $f_i$ is the field of $i$th trap in the donor site appeared in geometric moments of the trap fields
$S_n=\sum_{i=1}^{N} f_i^{n}$, $N$ is the number of contributing traps, $E_0$ is the electric field bias, and $\beta$ is the polarizability of the emitter. 

Our model qualitatively explains the stabilization dynamics of trap states by framing them as a competition between carrier capture and release rates. Based on this model, charge fluctuation is diminished when either rate dominates, pushing the occupation probability ($p$) toward the stable limits of 0 or 1. Under electrical suppression, band bending increases the tunneling rate \cite{ganichev2000distinction}, driving the occupation probability $p$ toward 0, illustrated in Fig.~\ref{fig:Fig3}(e). At higher fields, the bias-noise coupling term $2E_0^2S_2$ begins to dominate, leading to monotonic spectral broadening and thus defining a sweet spot for linewidth minimization, reflecting the behavior in Fig. \ref{fig:Fig3}(b).  In the case of optical suppression, above-band pumping increases the capture rate via photo-generated carriers \cite{PhysRev.87.835}, pushing $p$ toward 1, which is illustrated in Fig. \ref{fig:Fig3}(f). Increased occupation leads to a blue-shifted central wavelength and a narrower emission linewidth, as shown in Fig. \ref{fig:Fig3}(d). Additional detail about the extracted fit parameters, derivation and Monte Carlo simulation based on this model is provided in Supplemental Material section VI.

\section{Conclusion and outlook}
\label{sec:develop}

We show that laterally applied electric fields in ZnSe quantum wells enable local tuning of single impurity–bound exciton emission energies, compensating for spectral inhomogeneity arising from strain and charge environment imperfections. Beyond spectral control, these fields can suppress trap-induced charge fluctuations, resulting in observable linewidth narrowing for the donor-bound exciton transition. We also verify this through weak above-band optical pumping and a statistical model, qualitatively describing the observed phenomena.  Based on current results, the precise tuning and reduction of spectral wandering may lead to higher visibility in Hong–Ou–Mandel (HOM) from two separate devices, addressing previous challenges. Additionally, it can potentially lead to enhanced spin coherence times. Looking ahead, embedding the quantum well in a Schottky diode structure would introduce an out-of-plane field that directly addresses the vertical confinement axis, offering stronger control and the potential to restore transform-limited linewidths. Together, these observations establish electric fields as a versatile approach for both finely tuning emission energies and stabilizing the charge environment in impurity-bound excitonic emitters. 

\section*{Acknowledgments}
The authors acknowledge Shahriar Aghaei,
Kai-Mei Fu, Douglas L Irving and Kelsey J Mirrielees for valuable
discussions. The Waks group would like to acknowledge support from the AFOSR (grant \#FA95502010250, grant \#FA95502410266, and grant \#FA95502310667). The Pawlis group would
like to acknowledge support from the Deutsche Forschungsgemeinschaft (DFG 337456818) under Germany's Excellence Strategy - Cluster of Excellence Matter and Light for Quantum Computing  (ML4Q) EXC 2004/1-390534769. The funder played no role in study design, data collection, analysis and interpretation of data, or the writing of this manuscript.

\section*{DATA AND MATERIALS AVAILABILITY}
All of the data that support the findings of this study are reported in the main text and Supplementary Materials. Source data are available from the corresponding authors on reasonable request.

\section*{Competing interests}
The authors declare that they have no competing interests.

\section*{Author contributions} A.A. and E.W. conceived the experiment. C.F., N.D., and Y.K. fabricated the quantum well and performed initial characterization. A.A. patterned the electrodes and made the connections.  
A.A. performed the experiment. Y.J. and A.B. supported setting up the experiment. A.A. analyzed the data. A.A., J.B., and E.W. prepared the manuscript. All authors discussed the results and confirmed the manuscript. E.W. and A.P. supervised the experiment.

\bibliographystyle{ieeetr}   
\bibliography{refs.bib}          


\end{document}


\title{Supplemental Material for "Electrical Control of Optically Active Single Spin Qubits in ZnSe"}

\author{Amirehsan Alizadehherfati}
\email{herfati@umd.edu}
\affiliation{Institute for Research in Electronics and Applied Physics and Joint Quantum Institute, University of Maryland, College Park, Maryland 20742, USA}
\affiliation{Department of Electrical and Computer Engineering, University of Maryland, College Park, MD 20742, USA}

\author{Yuxi Jiang}
\affiliation{Institute for Research in Electronics and Applied Physics and Joint Quantum Institute, University of Maryland, College Park, Maryland 20742, USA}
\affiliation{Department of Electrical and Computer Engineering, University of Maryland, College Park, MD 20742, USA}

\author{Nils von den Driesch}
\affiliation{Peter-Gr\"unberg Institute (PGI-9 \& PGI-10), Forschungszentrum J\"ulich GmbH, 52428 J\"ulich, Germany}
\affiliation{JARA-Fundamentals of Future Information Technology, Forschungszentrum J\"ulich and RWTH Aachen University, 52062 Aachen, Germany}

\author{Christine Falter}
\affiliation{Peter-Gr\"unberg Institute (PGI-9 \& PGI-10), Forschungszentrum J\"ulich GmbH, 52428 J\"ulich, Germany}
\affiliation{JARA-Fundamentals of Future Information Technology, Forschungszentrum J\"ulich and RWTH Aachen University, 52062 Aachen, Germany}

\author{Yurii Kutovyi}
\affiliation{Peter-Gr\"unberg Institute (PGI-9 \& PGI-10), Forschungszentrum J\"ulich GmbH, 52428 J\"ulich, Germany}
\affiliation{JARA-Fundamentals of Future Information Technology, Forschungszentrum J\"ulich and RWTH Aachen University, 52062 Aachen, Germany}

\author{Jasvith Raj Basani}
\affiliation{Institute for Research in Electronics and Applied Physics and Joint Quantum Institute, University of Maryland, College Park, Maryland 20742, USA}
\affiliation{Department of Electrical and Computer Engineering, University of Maryland, College Park, MD 20742, USA}

\author{Amirehsan Boreiri}
\affiliation{Institute for Research in Electronics and Applied Physics and Joint Quantum Institute, University of Maryland, College Park, Maryland 20742, USA}
\affiliation{Department of Electrical and Computer Engineering, University of Maryland, College Park, MD 20742, USA}

\author{Alexander Pawlis}
\affiliation{Peter-Gr\"unberg Institute (PGI-9 \& PGI-10), Forschungszentrum J\"ulich GmbH, 52428 J\"ulich, Germany}
\affiliation{JARA-Fundamentals of Future Information Technology, Forschungszentrum J\"ulich and RWTH Aachen University, 52062 Aachen, Germany}

\author{Edo Waks}
\email{edowaks@umd.edu}
\affiliation{Institute for Research in Electronics and Applied Physics and Joint Quantum Institute, University of Maryland, College Park, Maryland 20742, USA}
\affiliation{Department of Electrical and Computer Engineering, University of Maryland, College Park, MD 20742, USA}


\maketitle
\onecolumngrid

\section*{Methods}\label{Methoods}
\subsection{Device description}\label{subsec:device}

The ZnMgSe/ZnSe/ZnMgSe quantum well heterostructure was grown by molecular-beam epitaxy on GaAs. The ZnSe QW is \(4.8\,\mathrm{nm}\) thick and is sandwiched between \(27.7\,\mathrm{nm}\) ZnMgSe barriers above a \(14\,\mathrm{nm}\) ZnSe buffer layer on GaAs.  Chlorine donors were incorporated into the QW by \(\delta\)-doping at a sheet density of \(\sim 10^{10}\,\mathrm{cm}^{-2}\). A 100 nm thick $\mathrm{AlO}_{\mathrm{x}}$ capping layer was grown, acting as a protective layer for ZnSe well and heat insulation between the electrodes and the quantum well. Metal electrodes were defined by a single-step lift-off process. After solvent cleaning and a dehydration bake, a \SI{300}{nm} layer of positive e-beam resist (ZEP-520A) was spin-coated and soft-baked. Electrode patterns (\(4\,\mu\mathrm{m}\) width) with gap spacings of \(1\)–\(6\,\mu\mathrm{m}\) were written by electron-beam lithography (Elionix ELS-G100) and developed in ZED-N50, rinsed in isopropanol, and briefly descummed in \(\mathrm{O}_2\) plasma to promote clean lift-off. Ti/Au (10/100\,nm) was deposited by electron-beam evaporation at base pressure \(<2\times10^{-7}\,\mathrm{Torr}\) with substrate rotation. Lift-off was performed in heated Remover~PG (\(\approx 60^{\circ}\mathrm{C}\)), followed by acetone and IPA rinses and \(\mathrm{N}_2\) dry. Contact pads were defined in the same step to facilitate silver paste bonding.

\subsection{Experimental Setup}\label{subsec:setup}
The sample is mounted in a closed loop cryostat (Bluefors) working at \SI{3.6}{K} equipped with a superconducting magnet providing fields up to 9 T in Voigt configuration (American Magnetics). We used a DC voltage supply (Keysight, E36106B) for applying bias to the electrodes. Electrical connections were made by silver-paste bonding of the device pads to the sample-holder PCB pins, with the chip mounted on the holder in direct thermal contact with the fridge cold finger. A free-space confocal microscope is used for excitation and collection, with a \(0.68\)-NA objective and a focal length of 3 mm. Above-band excitation at \(405\,\mathrm{nm}\) is provided by a diode laser (Thorlabs, LP405\textendash SF10). The resonant excitation and the laser scanning are performed using a tunable diode laser (TOPTICA, DL pro), a frequency-doubled pulsed
Ti:sapphire laser (Mira) and scanning Fabry-Perot interferometer (Thorlabs, SA30-47). We used a \SI{600}{MHz} resolution wavemeter (HighFinesse, Angstrom WS/6) to monitor the central wavelength of resonant laser. The resonant beam is delivered through polarization-maintaining single-mode fiber (PM-SMF) and then passes a linear polarizer, half-wave plate, and quarter-wave plate to prepare the incident polarization for cross-polarized detection. The collected signal passes through a different set of quarter-wave plate, half-wave plate, and linear polarizer, and is then collected by a polarization-maintained single mode fiber. Spectra are acquired with a Princeton Instruments spectrometer comprising a monochromator (1800\,g/mm grating) and a 1340 pixel CCD camera. For time-resolved measurements, the collected signal is detected on fiber-coupled superconducting single-photon detectors (QuantumOpus). Detector outputs are time-tagged by a HydraHarp400 (PicoQuant). We observe a response time (jitter) of about 
\SI{30}{\pico s} using 4 ps pulses.  

\section{PHOTOLUMINESCENCE MAGNETOSPECTROSCOPY}

\begin{figure}[h]
  \centering
  \includegraphics[width=\columnwidth]{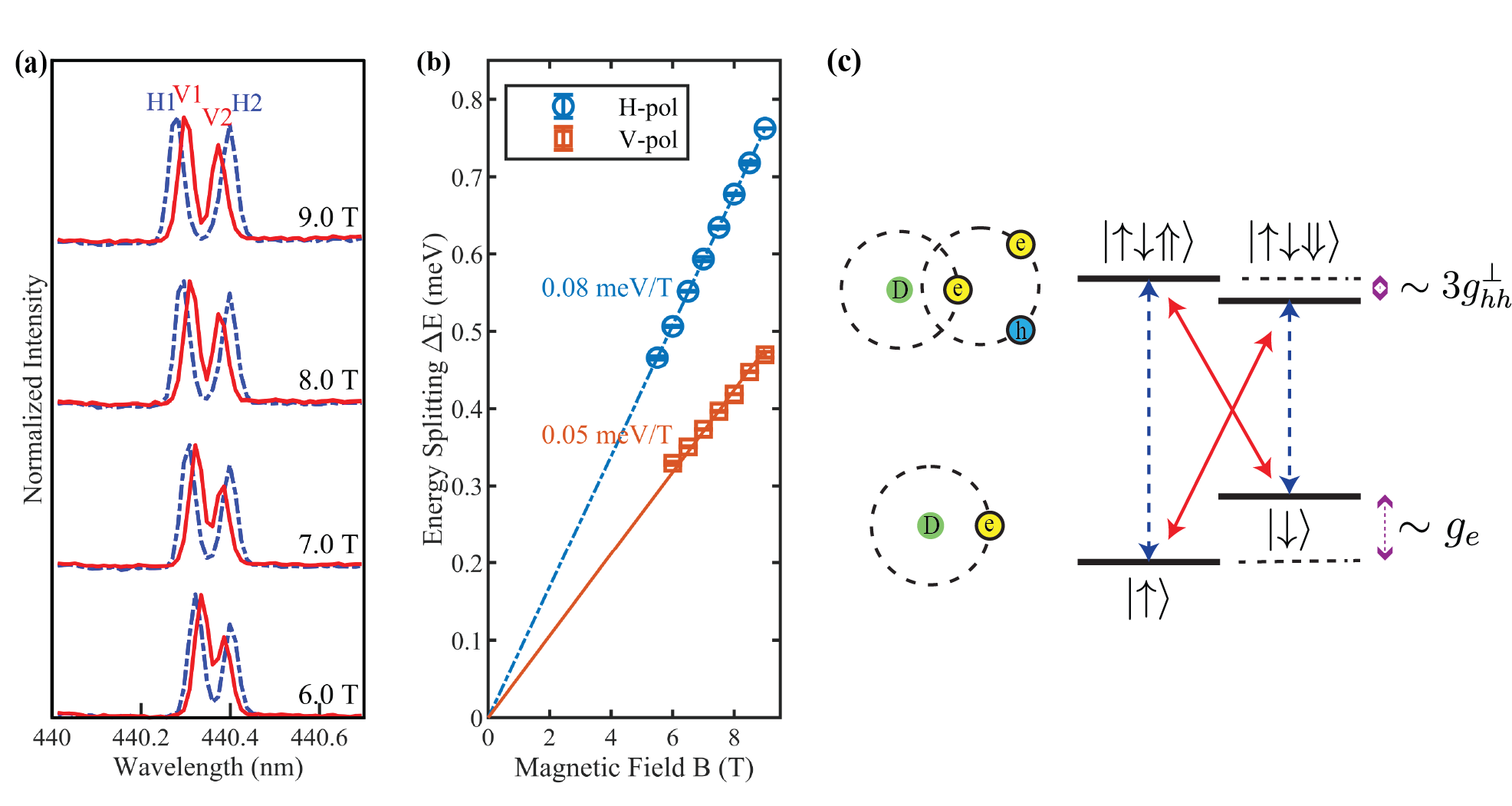} 
  \caption{Polarization-resolved magnetospectroscopy of donor-bound emission}
  \label{fig:S1}
\end{figure}

To probe the origin of the emitters, we perform polarization-resolved magnetospectroscopy in the Voigt geometry. Fig. \ref{fig:S1}a shows the photoluminescence emission spectra as a function of magnetic fields up to 9 T. The magnetic field lifts the degeneracy of the electron ($\ket{\uparrow}$ or $\ket{\downarrow}$) and hole  ($\ket{\Uparrow}$ or $\ket{\Downarrow}$) spin in ground and excited state, leading to splitting in each polarization. 

Fig. \ref{fig:S1}b shows the splitting between lines as a function of magnetic field, for each polarization. We extract each splitting using a double Voigt profile. We fit a solid line to the separation values as a function of magnetic fields. Based on this calculation, we extract the effective Zeeman splitting rate associated with each polarization.  We find $g_H = 1.463 \pm 0.002$ and $g_V = 0.915 \pm 0.014$. To extract the electron and hole g-factors, we analyze the level structure and Zeeman splitting associated with the optical transitions (Fig. \ref{fig:S1}c). For the donor-bound exciton, the splitting behavior corresponds to a ground-state electron and an excited-state heavy hole. The ground state splits according to the electron g-factor, while the excited state—formed by a negatively charged trion ($X^-$)—splits with the hole g-factor due to the singlet pairing of the two electrons. Based on the measured Zeeman splittings in horizontal (H) and vertical (V) polarizations, we extract the effective g-factors as
$g_e=(g_H + g_V)/2=1.189\pm 0.014$ for electron and $3g_{hh}=(g_H - g_V)/2=0.274\pm 0.014$ for heavy hole.

\section{Polarization Selectivity from DX line}

\begin{figure}[h]
  \centering
  \includegraphics[width=460 px]{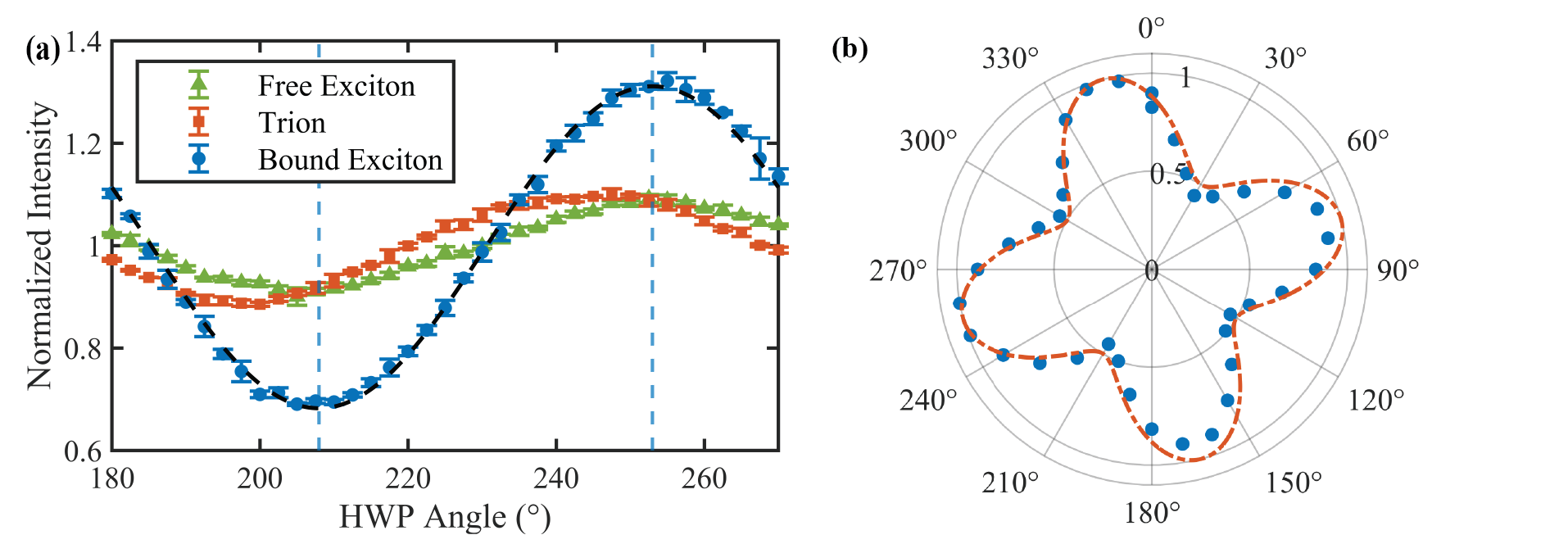}
  \caption{Polarization-resolved photoluminescence from donor-bound (DX), trion (X$^-$), and free-exciton (FX) transitions.  
  The DX shows the largest polarization visibility, consistent with a single dipole radiating in the field between two gold electrodes.}
  \label{fig:S2}
\end{figure}

Fig. \ref{fig:S2}a presents the normalized emission intensity of the impurity-bound exciton (DX), trion (X$^-$), and free exciton (FX) transitions as a function of half-wave-plate angle.  
All lines exhibit measurable modulation, but the DX transition shows the highest polarization visibility, indicating a stronger dipole moment, probably aligned to the local field direction between electrodes.  
The X$^-$ and FX transitions display weaker, yet nonzero modulation due to mixed selection rules and averaging over multiple dipole orientations.  
The dataset is fitted to
\begin{equation}
I(\theta)=I_0+I_1\cos\!\bigl[2(\theta-\theta_0)\bigr],\qquad
\mathcal{V}\equiv I_1/I_0,
\end{equation}
where $\mathcal{V}$ denotes the polarization visibility extracted from the sinusoidal amplitude ratio. For the donor line (Fig. \ref{fig:S2}b), we observe $\mathcal{V} = 0.4$.

We attribute this effect to the propagation of light between two metallic electrodes. Consider two parallel metallic faces at $x=\pm L/2$ (air in between) and an oscillating dipole at $x=0$. Let $k=2\pi/\lambda_0$ and define the single–pass phase across the gap
\begin{equation}
\phi \equiv kL=\frac{2\pi L}{\lambda_0}.
\end{equation}
Each face is characterized by a complex amplitude reflectivity $r=|r|e^{i\delta}$ at $\lambda_0$ (normal-incidence Fresnel coefficient), with $r=(N-1)/(N+1)$ for $N=n+ik$ \cite{HechtOptics}.  

We can decompose the in-plane radiation into eigen-axes relative to the faces. Electric field can be perpendicular ($\perp$) or parallel ($\parallel$) to the faces. For a conducting plane, the image-dipole has the same orientation for the normal component and the opposite orientation for the tangential component (parity result) \cite{NovotnyHecht2012}. Writing $R\equiv r\,e^{i\phi}$, the coherent field at the source plane is the direct term plus all multiply reflected terms:
\begin{equation}
\mathcal E_{\perp}=(1+R)\sum_{m=0}^{\infty}R^{2m}
=\frac{1+R}{1-R^{2}},\qquad
\mathcal E_{\parallel}=(1-R)\sum_{m=0}^{\infty}R^{2m}
=\frac{1-R}{1-R^{2}}.
\label{eq:Eseries}
\end{equation}
The analyzer-aligned intensities are $I_{\perp,\parallel}\propto|\mathcal E_{\perp,\parallel}|^2$, giving the visibility
\begin{equation}
\mathcal V(\phi,r)=
\frac{|\mathcal E_{\perp}|^2-|\mathcal E_{\parallel}|^2}
     {|\mathcal E_{\perp}|^2+|\mathcal E_{\parallel}|^2}
=\frac{2|r|\cos(\phi+\delta)}{1+|r|^2}.
\label{eq:Vclosed}
\end{equation}

Using Johnson–Christy gold data at $0.440~\mu$m\cite{JohnsonChristy1972,RIinfoAu}, one obtains
$N=n+ik\simeq 1.417+1.932\,i$, hence
$r=(N-1)/(N+1)$ with $|r|\simeq0.639$ and $\delta\simeq0.684~\text{rad}$.  
For $L=1.00~\mu$m, $\phi=2\pi L/\lambda_0=14.280~\text{rad}$; inserting into Eq.~(\ref{eq:Vclosed}) yields
\[
\ \mathcal V \approx -0.667\ \quad (|\mathcal V|\approx 0.667).
\]
In experiment the magnitude can be somewhat lower (e.g.\ $\sim0.3$–$0.5$) due to finite electrode height/edge roughness (reduced effective $|r|$), tiny phase detuning from nm-scale gap/wavelength errors, slight emitter off-centering, finite NA (angle-dependent $r$), and weak unpolarized background; all act to compress the ideal contrast without altering Eq.~(\ref{eq:Vclosed}).

\section{Excited state lifetime measurement}
\begin{figure}[h]
  \centering
  \includegraphics[width=4.5375 in]{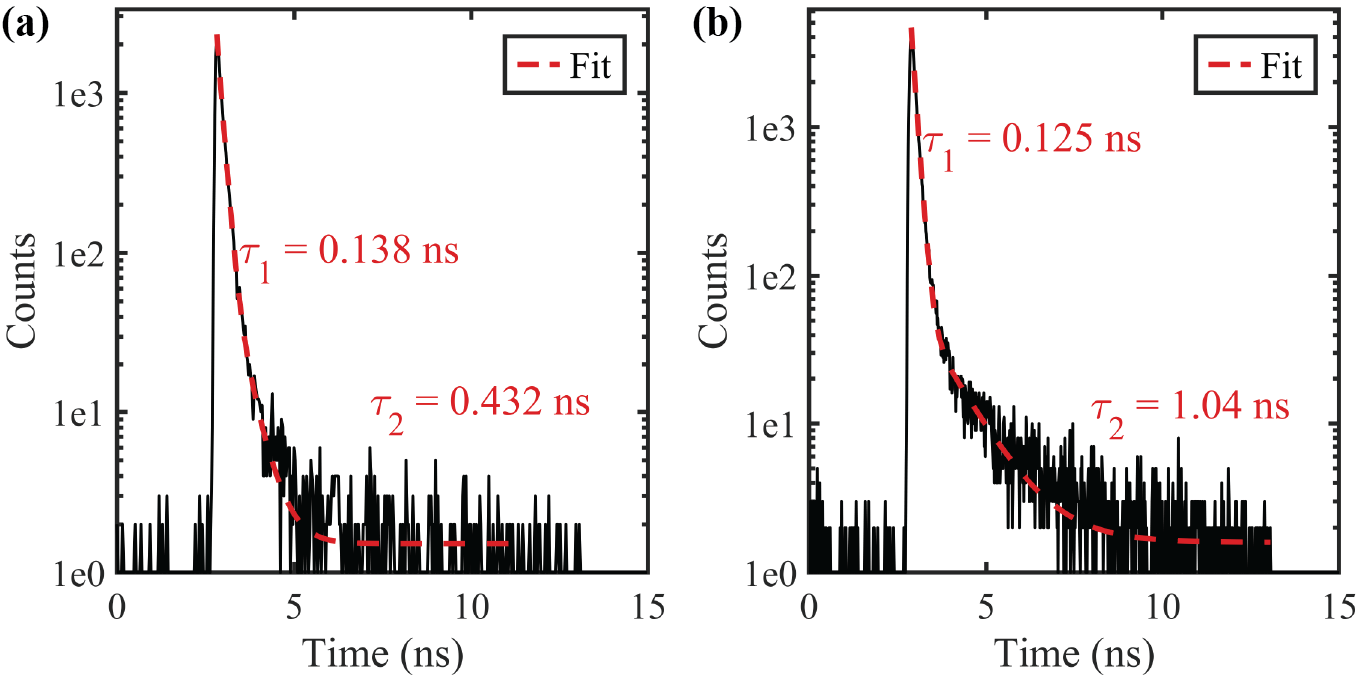}
  \caption{(a)(b) Time resolved measurements on the excited state lifetime of single donor bound exciton emissions, showing an average lifetime of $131\pm4~$ps for two measured sites.}
  \label{fig:S44}
\end{figure}

We characterize the lifetime of donor emissions in the regions between the electrodes. Fig. \ref{fig:S44} shows the excited state double exponential decay extracted from a pulsed above-band excitation measurement. We did not observe any discernible effects from the electrode structure. Additionally, in comparison to previous bulk studies on the same system \cite{Karasahin2022}, we observe a shorter excited-state lifetime, indicating an increased radiative rate. This effect is likely attributable to advancements in growth quality. 

\section{Electrical Tuning based on Lateral Fields}

\subsection{Finite Element Method Simulation}
To determine the polynomial coefficients describing the dependence of the emitter transition frequency on the applied bias, we performed finite-element electrostatic simulations (COMSOL Multiphysics). The model reproduces the device stack—coplanar metal electrodes atop a continuous oxide dielectric over a ZnSe quantum well (on GaAs)—with layer thicknesses and electrode spacing taken from the fabricated structure. We solved the static Laplace problem with the electrodes held at ($\pm V$) and all other boundaries electrically insulating, and we extracted the lateral electric-field magnitude at the emitter site, evaluated at the mid-plane of the ZnSe well coincident with the ($\delta$)-doped layer. These fields provided the voltage-to-field conversion used to fit the Stark-shift polynomial coefficients.

\begin{figure}[h]
  \centering
  \includegraphics[width=405 px]{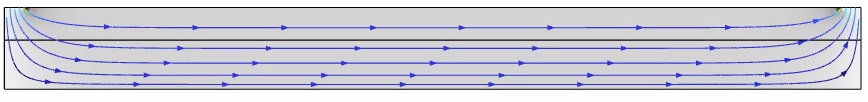}
  \caption{Electric Field Simulation considering oxide top layer and ZnSe well}
  \label{fig:comsol}
\end{figure}

Figure \ref{fig:comsol} plots the simulation results, showing that the field in the ZnSe well is predominantly \emph{lateral}. In the thin–stack, coplanar–electrode limit ($t_{\mathrm{ox}}+t_{\mathrm{ZnSe}}\ll L$), the tangential field is continuous across interfaces, so to leading order the lateral field inside ZnSe equals the macroscopic in–gap field,
\begin{equation}
F_{\parallel}^{\mathrm{(ZnSe)}} \;\simeq\; \frac{V_{\mathrm{bias}}}{L}.
\label{eq:lateral-leading}
\end{equation}
For our sample device ($L=1~\mu$m), Eq.~\eqref{eq:lateral-leading} gives $F_{\parallel}=10~\mathrm{V}/1~\mu\mathrm{m}=100~\mathrm{kV/cm}$ at $V_{\mathrm{bias}}=10$~V, consistent with the COMSOL value $91.1~\mathrm{kV/cm}$ at the donor site. We therefore use a single geometry factor $\eta\simeq0.911$ to account for weak fringing and depth dependence:
\begin{equation}
F_{\mathrm{ext}}(V) \;=\; \eta\,\frac{V}{L}, 
\qquad \eta=0.911,\;\;L=1~\mu\mathrm{m},
\label{eq:external-field}
\end{equation}
so $1$~V $\mapsto$ $9.11~\mathrm{kV/cm}$ (external field) at the emitter.

Because the emitter is embedded in a dielectric, the microscopic field felt by its dipole differs from $F_{\mathrm{ext}}$. Using the Lorentz (virtual–cavity) local–field model,
\begin{equation}
F_{\mathrm{loc}} \;=\; \frac{\varepsilon_r+2}{3}\;F_{\mathrm{ext}},
\label{eq:lorentz}
\end{equation}
with ZnSe $\varepsilon_r\simeq 8.8$, yielding a factor $(\varepsilon_r+2)/3 \approx 3.6$. Combining Eqs.~\eqref{eq:external-field}–\eqref{eq:lorentz},
\begin{equation}
F_{\mathrm{loc}}(V) \;=\; \frac{\varepsilon_r+2}{3}\,\eta\,\frac{V}{L}
\;\approx\; 3.6\times 0.911 \times \frac{V}{L}
\;=\; 32.8\,\frac{V}{\mathrm{V}} \frac{\mu\mathrm{m}}{L}~\mathrm{kV/cm}.
\label{eq:VtoFloc}
\end{equation}
Numerically, $V_{\mathrm{bias}}=10$~V gives $F_{\mathrm{ext}}\approx 91.1~\mathrm{kV/cm}$ and $F_{\mathrm{loc}}\approx 3.28\times 10^{2}~\mathrm{kV/cm}$. This $F_{\mathrm{loc}}(V)$ is the field value used to extract the Stark coefficients reported in the main text, for the array with $L = 2~\mu\mathrm{m}$. 

\subsection{Spectral Tuning of Free Exciton and Trion}

\begin{figure}[h]
  \centering
  \includegraphics[width=405 px]{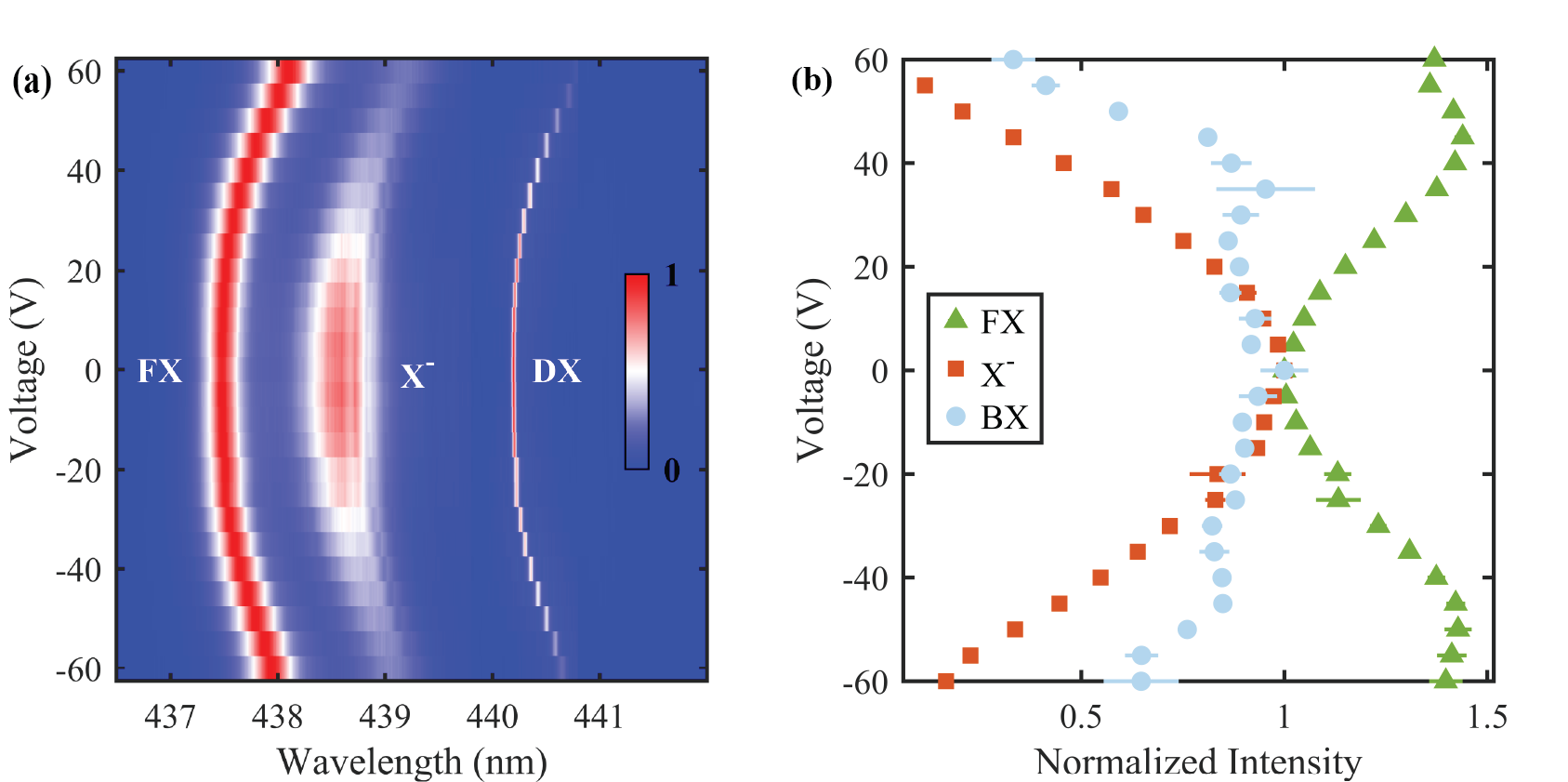}
  \caption{Spectral response of spectrum components to electric bias}
  \label{fig:S3}
\end{figure}

In the main text, we focus on the spectral tuning behavior of impurity-bound excitons. A similar analysis can be extended to the free exciton and trion transitions. Fig. \ref{fig:S3}a presents a color map of all three components as a function of applied voltage. 

The tuning slopes for the free exciton, trion, and bound exciton are comparable, indicating that the Stark shift affects all three transitions  . This suggests that the exciton binding energies in these systems are modified by the lateral electric field via polarization effects across the quantum well.
Despite their similar tuning rates, the three species exhibit distinct stability profiles under the applied field, as reflected in their respective emission intensities. As shown in Fig. \ref{fig:S3}b, the trion intensity decreases monotonically with increasing voltage. This behavior is in contrast to the bound exciton, which maintains stable emission under the same conditions. Although both species share structural similarities, this divergence highlights the enhanced localization and robustness of the bound exciton compared to the more delocalized trion. As the trion population decreases, the probability of observing free exciton emission increases—consistent with field-assisted dissociation of the trion complex.

\section{power dependent Photoluminescence spectroscopy}

\begin{figure}[h]
  \centering
  \includegraphics[width=450 px]{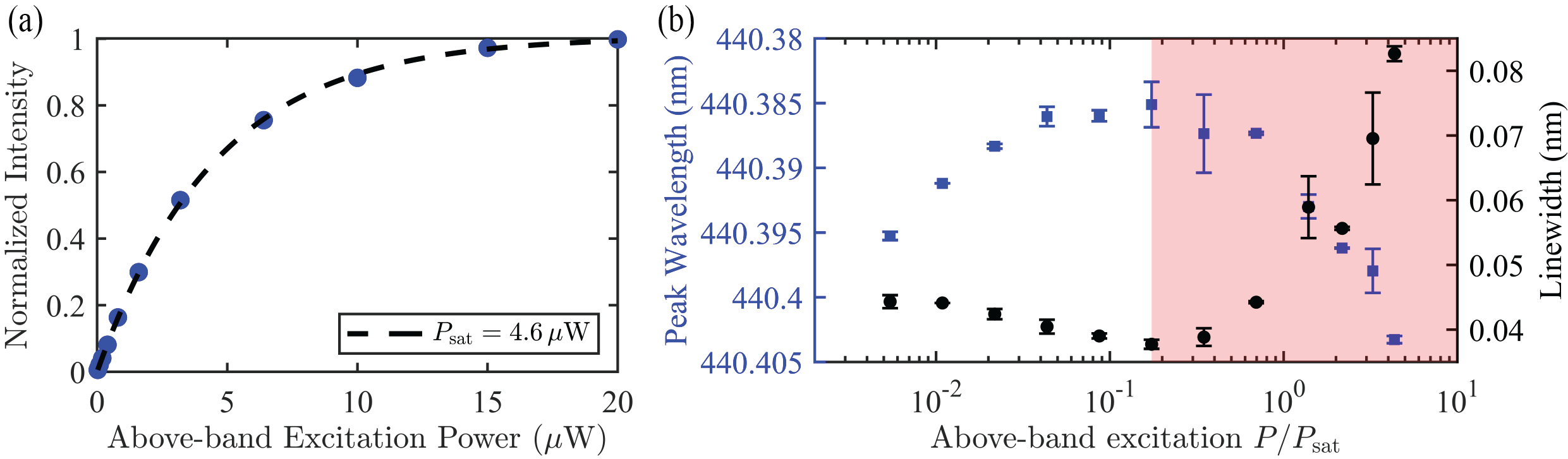}
  \caption{Power dependence measurement. (a) Emission as a function of above-band excitation power (b) Dynamics of central wavelength and linewidth of bound-exciton emission as a function of above-band excitation power.}
  \label{fig:S4}
\end{figure}

Figure \ref{fig:S4}a shows the emission intensity of the bound-exciton line as a function of above-band excitation power. The data exhibit clear saturation at \SI{4.6}{\micro W}, in agreement with previous studies on bulk ZnSe single emitters. We extract this saturation power by fitting the two-level saturation relation 
\begin{equation}
    I(P) = I_{\rm sat}\frac{P}{P+ P_{\rm sat}}
\end{equation}
to the experimental data, showing as dashed curve. This agreement confirms that the system behaves as an isolated two-level emitter.

Figure \ref{fig:S4}b shows the linewidth and central wavelength of the emission versus normalized power $P/P_{\rm sat}$. We extract both values from a Voigt profile fit to the experimental data. The linewidth narrows up to a turnover at $P\approx 0.17 P_{\rm sat}$, beyond which it broadens. Similarly, the peak wavelength blueshifts up to that same point and then redshifts at higher powers. Below this threshold, we attribute the narrowing and blueshift to trap-state filling, which reduces spectral diffusion and suppresses the built-in DC Stark shift by neutralizing charge traps. Above this threshold, power broadening dominates, leading to linewidth broadening and a redshift. This behavior is critical for photoluminescence excitation experiments, as it clarifies the role of above-band pumping and underscores the need to isolate purely electric-field–induced effects.

\section{Stabilization of Local Electric Field}

\subsection{Theory of Trap State Mediated Broadening}

We consider a uniform distribution of traps in the vicinity of donor site located at $\mathbf r_0$. We index the traps by $i=1,\dots,N$, where trap $i$ is located at $r_i$ in the same coordinate system as donor. Each trap toggles between empty/fill occupancy, behaving as a two-state Markovian telegraph process $s_i(t)\in\{0,1\}$ 
\begin{align}
 0 \xrightleftharpoons[k_i^-]{\;\;k_i^+\;\;} 1.
\end{align}
where $k_i^{\pm}$ is the rate of capture/release. The dynamics of each trap can be written as a rate equation for the probability of being occupied, $P_1^{(i)}(t)=\Pr[s_i(t)=1]$.

\begin{align}
 \frac{d}{dt} P_1^{(i)}(t) = k_i^+ \, P_0^{(i)}(t) - k_i^- \, P_1^{(i)}(t),
 \qquad P_0^{(i)}=1-P_1^{(i)}.
\end{align}
 In steady state ($dP_1^{(i)}/dt=0$),
\begin{align}
  \mathbb{E}[s_i]=p_i=\frac{k_i^+}{k_i^++k_i^-}, \qquad
 \mathrm{Var}[s_i]= p_i(1- p_i), \qquad \tau_i=\frac{1}{k_i^++k_i^-}\ .
 \label{eq:ss-n}
\end{align}

In the presence of charge in the trap $s_i = 1$, it produces a Coulomb field at the emitter site
\begin{align}
 \bm f_i
 = \frac{e}{4\pi \varepsilon}\, \frac{\mathbf r_0-\mathbf r_i}{\abs{\mathbf r_0-\mathbf r_i}^3},
 \label{eq:Ei-vector}
\end{align}
where $\varepsilon=\varepsilon_0\varepsilon_r$ is the host ZnSe permittivity. The total field which is the superposition of all trap fields can be written as  
\begin{equation}
\bm E_{\text{tot}}(t)=\bm E_0+\sum_{i=1}^N s_i(t)\,\bm f_i,
\end{equation}
where $\bm E_0$ is the bias field. Based on the applied bias results, we observe a dominant quadratic term from the emitter. Therefore, we can write the frequency response of the emitter as
\begin{equation}
\Delta\omega(t)=\beta\,\|\bm E_{\text{tot}}(t)\|^2.
\label{eq:dw_bias_start}
\end{equation}
with quadratic Stark coefficient $\beta$. Expanding the square and defining
\begin{equation}
u_i\equiv 2\,\bm E_0\!\cdot\!\bm f_i = 2E_0f_i\cos\phi_i,\qquad
a_i\equiv \|\bm f_i\|^2=f_i^2,\qquad
b_{ij}\equiv 2\,\bm f_i\!\cdot\!\bm f_j=2 f_i f_j \cos\phi_{ij}\ (i\neq j),
\label{eq:uab_defs_bias}
\end{equation}
we obtain
\begin{equation}
\Delta\omega(t)=\beta\Big[E_0^2+\sum_{i=1}^N (u_i+a_i)\,s_i(t)+\sum_{1\le i<j\le N} b_{ij}\,s_i(t) s_j(t)\Big].
\label{eq:dw_bias_expand}
\end{equation}
To find a simplified closed form solution, we make the following assumptions.
\begin{enumerate}[label=(\roman*)]
    \item Traps are identical, meaning that the capture/release rate $k^\pm_i = k^\pm$ is equal among them.
    \item Traps are independent, meaning that $\left\lbrace s_i(t)\right\rbrace$ are independent.
    \item Traps are isotropically distributed in the plane of quantum well, surrounding the donor.
    \item Optical coherence $T_2$ of the transition is much shorter than trap switching time $\tau$ and spectrometer integration $T$ is much longer than $\tau$. 
\end{enumerate}
Therefore, each recorded spectrum is the time average over the stationary distribution of $\Delta\omega(t)$. 
 We need to extract the average and variance of $\Delta\omega(t)$, which indicates the central wavelength and observed gaussian broadening.

\subsubsection{Average of Frequency}
Considering the independence of $\{s_i(t)\}$, we can write $\mathbb{E}[s_i(t)]=p$ and, for $i\neq j$, $\mathbb{E}[s_i(t) s_j(t)]=p^2$. Taking expectations of Eq.~\eqref{eq:dw_bias_expand} over time and angle gives
\begin{equation}
\mu_\omega\equiv\mathbb{E}[\Delta\omega]
= \beta\Big[E_0^2+p\sum_i a_i+p\sum_i\langle u_i\rangle+p^2\sum_{i<j}\langle b_{ij}\rangle\Big]=  \boxed{\,\beta\Big(E_0^2+p\,S_2\Big)\,}\ 
,\qquad S_2\equiv\sum_{i=1}^N f_i^2.
\label{eq:mu_bias_final}
\end{equation}
where we used the isotropy assumptions  Eq.~\eqref{eq:angle_u} and Eq.~\eqref{eq:angle_b}. Thus the bias adds a deterministic shift $\beta E_0^2$; the trap contribution to the mean is unchanged from zero bias.

\subsubsection{Variance of Frequency}
We can separate the term to linear and pair according to $\{s_i(t)\}$
\begin{equation}
L\equiv\sum_i (u_i+a_i)\,s_i(t),\qquad Y\equiv\sum_{i<j} b_{ij}\,s_i(t) s_j(t),
\end{equation}
so that
\begin{equation}
\sigma_\omega^2=\mathrm{Var}[\Delta\omega]
=\beta^2\Big(\mathrm{Var}L+\mathrm{Var}Y+2\,\mathrm{Cov}(L,Y)\Big).
\label{eq:var_bias_split}
\end{equation}
Since $\mathrm{Var}[s_i(t)]=p(1-p)$ and the $\{s_i(t)\}$ are independent, we can write the linear part as
\begin{equation}
\mathrm{Var}L=\sum_i (u_i+a_i)^2\,\mathrm{Var}[s_i(t)]=p(1-p)\sum_i (u_i+a_i)^2.
\end{equation}
The isotropic approximation with Eq.~\eqref{eq:angle_u2} gives the following.
\begin{equation}
\big\langle\mathrm{Var}L\big\rangle=p(1-p)\sum_i\big(a_i^2+2E_0^2 a_i\big)
=p(1-p)\Big(S_4+2E_0^2 S_2\Big)\ ,\qquad S_4\equiv\sum_{i=1}^N f_i^4.
\label{eq:varL_bias_final}
\end{equation}
For the pair part, let $t_{ij}=b_{ij}s_i s_j$. Using $s_i^2=s_i$, the standard identity for a finite sum gives
\begin{equation}
\mathrm{Var}\,Y=\sum_{i<j}\mathrm{Var}(t_{ij})
+2\!\!\!\sum_{\substack{(i<j)<(k<l)}}\!\!\!\mathrm{Cov}(t_{ij},t_{kl}).
\label{eq:varY_cov_form}
\end{equation}
Because $s_i s_j$ is Bernoulli with success probability $p^2$,
\begin{equation}
\mathrm{Var}(t_{ij})=b_{ij}^2\,\mathrm{Var}(s_i s_j)
=b_{ij}^2\,p^2(1-p^2)
\label{eq:var_tij}
\end{equation}
Covariances vanish for disjoint pairs (all indices distinct) by independence:
\[
\mathrm{Cov}(t_{ij},t_{kl})=0 \quad \text{if}\quad \{i,j\}\cap\{k,l\}=\varnothing.
\]
For overlapping pairs that share exactly one index, e.g.\ $(ij)$ with $(ik)$:
\begin{align}
\mathrm{Cov}(t_{ij},t_{ik})
&= \mathbb{E}[t_{ij}t_{ik}] - \mathbb{E}[t_{ij}]\,\mathbb{E}[t_{ik}] \nonumber\\
&= b_{ij}b_{ik}\Big(\mathbb{E}[s_i s_j s_i s_k]-\mathbb{E}[s_i s_j]\,\mathbb{E}[s_i s_k]\Big) \nonumber\\
&= b_{ij}b_{ik}\Big(\mathbb{E}[s_i s_j s_k]-\mathbb{E}[s_i]^2\mathbb{E}[s_j]\mathbb{E}[s_k]\Big) \nonumber\\
&= b_{ij}b_{ik}\,p^3(1-p). \label{eq:cov_overlap_1}
\end{align}
Similarly,
\begin{equation}
\mathrm{Cov}(t_{ij},t_{jk})=b_{ij}b_{jk}\,p^3(1-p),\qquad
\mathrm{Cov}(t_{ik},t_{jk})=b_{ik}b_{jk}\,p^3(1-p).
\label{eq:cov_overlap_23}
\end{equation}
Hence
\begin{equation}
\mathrm{Var}\,Y
=\sum_{i<j} b_{ij}^2\,p^2(1-p^2)
+2\!\!\sum_{i<j<k}\!\!\Big(b_{ij}b_{ik}+b_{ij}b_{jk}+b_{ik}b_{jk}\Big)\,p^3(1-p).
\label{eq:varY_overlap_full}
\end{equation} 
Based on the isotropic approximation with Eq.~\eqref{eq:angle_b2} and Eq.~\eqref{eq:angle_mixed_zero}
\begin{equation}
\big\langle\mathrm{Var}Y\big\rangle
=\sum_{i<j}\langle b_{ij}^2\rangle\,p^2(1-p^2)
=\sum_{i<j} 2 f_i^2 f_j^2\,\big(p^2-p^4\big)
=p^2(1-p^2)\Big(S_2^2-S_4\Big).
\label{eq:varY_bias_final}
\end{equation}
Finally, every term in $\mathrm{Cov}(L,Y)$ contains one factor $b_{ij}$ multiplied by either $a_k$ or $u_k$; isotropy gives $\langle b_{ij}\rangle=\langle u_k b_{ij}\rangle=0$, hence
\begin{equation}
\big\langle\mathrm{Cov}(L,Y)\big\rangle=0.
\label{eq:covLY_bias_zero}
\end{equation}
Collecting \eqref{eq:varL_bias_final},\eqref{eq:varY_bias_final} and \eqref{eq:covLY_bias_zero},
\begin{equation}
\boxed{\
\sigma_\omega^2
=\beta^2\left[
p(1-p)\big(S_4+2E_0^2 S_2\big)\;+\;p^2(1-p^2)\big(S_2^2-S_4\big)
\right].\
}
\label{eq:sigma_bias_final}
\end{equation}
Compared to zero bias, the variance acquires an additive term $2\beta^2 E_0^2\,p(1-p)\,S_2$ originating from the cross term $2\bm E_b\!\cdot\!\sum_i s_i\bm f_i$ in \eqref{eq:dw_bias_expand}; physically, the bias couples to shot noise in trap occupancy. 

\subsubsection{Gaussian kernel and Voigt profile}

With $T\gg \tau$ the trap kernel is Gaussian,
\begin{equation}
I_G(\omega)=\frac{1}{\sqrt{2\pi}\,\sigma_\omega}\,
\exp\!\left[-\frac{(\omega-\omega_0-\mu_\omega)^2}{2\sigma_\omega^2}\right],
\label{eq:gaussian}
\end{equation}
with $\mu_\omega$ and $\sigma_\omega^2$ from Eqs.~\eqref{eq:mu_bias_final} and \eqref{eq:sigma_bias_final}.

Let $L_0(\omega)$ denote the homogeneous Lorentzian in the absence of traps,
\begin{equation}
L_0(\omega)=\frac{1}{\pi}\,\frac{\gamma}{(\omega-\omega_c)^2+\gamma^2},\qquad \gamma=\tfrac12\,\mathrm{FWHM}_L\ \ (\mathrm{FWHM}_L\simeq \tfrac{1}{\pi T_2}).
\label{eq:lorentzian}
\end{equation}
In the regime $T_2\ll \tau \ll T$, the measured line is the time/ensemble average
\begin{equation}
I(\omega)=\int_{-\infty}^{\infty} P(\Delta)\,L_0\!\big(\omega-\omega_0-\Delta\big)\,d\Delta
=\big[L_0 * \mathcal{N}(\mu_\omega,\sigma_\omega^2)\big](\omega-\omega_0),
\label{eq:conv_def}
\end{equation}
i.e.\ a Voigt profile. In closed form,
\begin{equation}
I(\omega)=\frac{1}{\sigma_\omega\sqrt{2\pi}}\,
\mathrm{Re}\!\left\{\,w\!\left(z\right)\right\},\qquad
z=\frac{\omega-\omega_0-\mu_\omega+i\gamma}{\sigma_\omega\sqrt{2}},
\label{eq:voigt_faddeeva}
\end{equation}
where $w(z)$ is the Faddeeva function. A convenient FWHM approximation is \cite{olivero1977empirical}
\begin{equation}
\mathrm{FWHM}_V \approx 0.5346\,\mathrm{FWHM}_L
+\sqrt{0.2166\,\mathrm{FWHM}_L^2+\mathrm{FWHM}_G^2},
\label{eq:voigt_fwhm}
\end{equation}
with $\mathrm{FWHM}_G=2\sqrt{2\ln 2}\,\sigma_\omega$.



\subsection{Optical suppression of trap fluctuations}

We consider an above-band pump of optical power $P$ that generates carriers at $G(P)=\eta P/(\hbar\omega)$ within the excited area $A$ and active thickness $d$. In the unsaturated (linear) CW regime the steady-state free-carrier density is
\begin{equation}
n(P)\simeq \frac{G(P)\tau_r}{A d}
=\frac{\eta\,\tau_r}{\hbar\omega\,A d}\,P \equiv \kappa P,
\label{eq:n_of_P}
\end{equation}
with recombination time $\tau_r$ and absorption/yield $\eta$; at cryogenic temperature the dark background is negligible once the pump is on. Trap charging (capture) follows Shockley–Read–Hall kinetics \cite{PhysRev.87.835}, giving a linear-in-$P$ rate
\begin{equation}
k^+(P)=k^+_0+\alpha_c P,\qquad
\alpha_c=\sigma_c v_{\rm th}\kappa,
\label{eq:kplus_linear}
\end{equation}
where $\sigma_c$ is the capture cross section and $v_{\rm th}$ the carrier velocity. Detrapping is weak at low $T$ but acquires a small pump dependence via photoionization \cite{lucovsky1965photoionization} and carrier-assisted neutralization; both are linear at low density, so
\begin{equation}
k^-(P)=k^-_0+\alpha_r P,\qquad
\alpha_r=\sigma_{\rm ph}\,\tilde{\kappa}+\alpha_1\,\kappa,\quad
\tilde{\kappa}=\frac{1}{\hbar\omega A},
\label{eq:kminus_linear}
\end{equation}
with $\sigma_{\rm ph}$ the photoionization cross section and $\alpha_1$ a carrier-assisted coefficient. Empirically $\alpha_r\ll\alpha_c$ over the modest-$P$ range of interest. The steady occupancy and switching time are then
\begin{equation}
p(P)=\frac{k^+_0+\alpha_c P}{k^+_0+k^-_0+(\alpha_c+\alpha_r)P},\qquad
\tau(P)=\frac{1}{k^+_0+k^-_0+(\alpha_c+\alpha_r)P}.
\label{eq:p_tau_linear}
\end{equation}
Equivalently, we write the steady occupancy in a saturating form
\begin{equation}
p(P)=p_0+\big(p_\infty-p_0\big)\frac{P}{P+P_{\rm sat}},
\label{eq:pPsat_form}
\end{equation}
which is algebraically identical to Eq.~\eqref{eq:p_tau_linear} under
\begin{equation}
p_0=\frac{k_0^+}{k_0^++k_0^-},\qquad
p_\infty=\frac{\alpha_c}{\alpha_c+\alpha_r},\qquad
P_{\rm sat}=\frac{k_0^+ + k_0^-}{\alpha_c+\alpha_r}.
\label{eq:pPsat_map}
\end{equation}
Based on the Eqs.~\eqref{eq:mu_bias_final} and \eqref{eq:sigma_bias_final}, we can write the mean and variance, in the absence of bias field as 
\begin{align}
\mu_\omega(P)&=\beta\,S_2\,p(P),
\label{eq:centerP_final}
\\
\sigma_\omega^2(P)&=\beta^2\Big[
p(P)\big(1-p(P)\big)\,S_4
+p^2(P)\big(1-p^2(P)\big)\,\big(S_2^2-S_4\big)
\Big].
\label{eq:sigmaP_final}
\end{align}
Increasing $P$ drives $p\to1$, so both $p(1-p)$ and $p^2(1-p^2)$ decrease monotonically. Therefore, the Gaussian component narrows while the center blue-shifts via Eq.~\eqref{eq:centerP_final}.

It is also useful to write this equation in a more suitable way for fit purposes. Starting from Eq.~\eqref{eq:sigmaP_final}, factoring out $(\beta S_2)^2$ and dividing inside the brackets by $S_2^2$:
\begin{equation}
\sigma_\omega^2(P)
=(\beta S_2)^2\!\left[
\frac{S_4}{S_2^2}\,p(P)\big(1-p(P)\big)
+\Big(1-\frac{S_4}{S_2^2}\Big)\,p^2(P)\big(1-p^2(P)\big)
\right].
\label{eq:sigmaP_factor}
\end{equation}
Without loss of generatility, we can define:
\begin{equation}
\hat{\kappa}\equiv\frac{S_4}{S_2^2}
\label{eq:kappahat_def}
\end{equation}
so Eq.~\eqref{eq:sigmaP_factor} becomes
\begin{equation}
\sigma_\omega^2(P)
=(\beta S_2)^2\!\left[
\hat{\kappa}\,p(P)\big(1-p(P)\big)
+\big(1-\hat{\kappa}\big)\,p^2(P)\big(1-p^2(P)\big)
\right].
\label{eq:sigmaP_khat}
\end{equation}
It is trivial to show that 
\begin{equation}
\frac{1}{N} \;\le\; \widehat{\kappa} \;\le\; 1.
\label{eq:kappa_bound}
\end{equation}
Using the isotropic and independent-trap approximation, together with a uniform area distribution of traps in the annulus, Eq. \ref{eq:kappahat_def} simplifies to  
\begin{equation}
\widehat{\kappa} \;\approx\; \frac{1}{3N}\,\bigl(a^{2}+1+a^{-2}\bigr),
\qquad a = \frac{r_{\max}}{r_{\min}},
\label{eq:kappa_kN}
\end{equation}

\subsection{Electrical suppression of trap fluctuations}

At $\sim\!4$\,K, detrapping under a static in-plane bias $\mathbf{E}_b$ proceeds athermally via field-assisted tunneling from a localized trap to the extended band. For a bound level of depth $\Phi$, and effective mass $m^\ast$, the potential along the escape path is well approximated by a triangular barrier in a uniform local field $E_0=\|\mathbf{E}_b\|$. The WKB action is
\begin{equation}
S(E_0)=\frac{2}{\hbar}\!\int_0^{x_t}\!\sqrt{2m^\ast\big(\Phi-qE_0 x\big)}\,dx
=\frac{4\sqrt{2m^\ast}}{3\hbar q}\,\frac{\Phi^{3/2}}{E_0}
\equiv \frac{E_\ast}{E_0},
\label{eq:WKB_action}
\end{equation}
where $x_t=\Phi/(qE_0)$ is the classical turning point and $E_\ast=\tfrac{4\sqrt{2m^\ast}}{3\hbar q}\,\Phi^{3/2}$ is the trap characteristic field. As a physical model for escape rate, we can use generalized Fowler–Nordheim form \cite{121690,SCHENK19921585,ganichev2000distinction,ganichev1997deep,fowler1928electron}
\begin{equation}
k^-_{\rm tun}(E_0)=\nu_t\,\mathcal{P}(E_0)\,\exp\!\left[-\left(\frac{E_\ast}{E_0}\right)^\gamma~\right],
\label{eq:FN_core}
\end{equation}
with an attempt frequency $\nu_t$ set by the bound-state level spacing, $\gamma$ an stretching coefficient, and a weak algebraic prefactor $\mathcal{P}(E_0)\propto E_0^{\alpha}$ that arises from barrier-shape and density-of-states factors. For deep traps and moderate fields one typically finds $\alpha\simeq 0$–$1$ and $\gamma\simeq 1$. Including a tiny zero-field baseline $k^-_0$ (residual tunneling through a thick barrier or very weak thermal emission), we adopt the fitting form
\begin{equation}
k^-(E_0)=k^-_0+B_0\,E_0^{\alpha}\exp\!\left[-\left(\frac{E_\ast}{E_0}\right)^\gamma~\right],
\label{eq:FN_full}
\end{equation}
where $B_0$ absorbs $\nu_t$ and geometric constants. This generalized model can phenomenologically captures the deviation from ideal Fowler-Nordheim form.

In the absence of injected carriers (no above-band light), the capture rate is set by dark background and we take, to leading order,
\begin{equation}
k^+(E_0)\approx k^+_0.
\label{eq:kplus_const}
\end{equation}
The occupancy thus decreases with field according to

\begin{align}
p(E_0)&=\frac{k^+_0}{k^+_0+k^-_0+B_0\,E_0^{\alpha}\exp\left[-\left(E_\ast/E_0\right)^\gamma~\right]},\\
&= \frac{p_0}{1+B\,E_0^{\alpha}\exp\left[-\left(E_\ast/E_0\right)^\gamma~\right]}
\label{eq:p_of_E}
\end{align}

where $B = B_0/p_0$.
Based on the Eqs.~\eqref{eq:mu_bias_final} and \eqref{eq:sigma_bias_final}, we can write the mean and variance as a function of field as
\begin{align}
\mu_\omega(E_0)&=\beta\big(E_0^2+p(E_0)\,S_2\big),
\label{eq:mu_with_Erates}\\
\sigma_\omega^2(E_0)&=\beta^2\!\left[
p(E_0)(1-p(E_0))\big(S_4+2E_0^2 S_2\big)
+p^2(E_0)(1-p^2(E_0))\big(S_2^2-S_4\big)
\right].
\label{eq:sigma_with_Erates}
\end{align}
As $E_0$ increases, enhanced tunneling lowers $p$ and reduces the trap-induced contribution to the mean shift, while the Gaussian width exhibits the expected nonmonotonic dependence on $p(E_0)$: starting from large $p$, the width can grow as $p$ approaches the maxima near $1/\sqrt{2}$ and $1/2$, and then narrows as traps empty further. 

\subsection{Monte Carlo Simulations}
To verify the validity of the closed-form solution we run Monte Carlo simulations with randomly placed electron traps near the donor. We draw \(N\) trap positions independently and uniformly in the annulus \((r_{\min}, r_{\max})\) and choose trap angles uniformly on \([0,2\pi)\) in the donor plane. Each trap follows a two-state Markov process with occupation probability \(p\). For comparison with the analytical model we generate \(n_{\mathrm{geom}}\) independent trap-geometry realizations, and for each geometry we take \(n_{\mathrm{snap}}\) temporal snapshots of the dynamics. For each \(p\) we form the wavelength distribution of the emitter and extract its mean and standard deviation. The mean gives the static shift of the central wavelength, and the standard deviation contributes to the observed linewidth. In what follows we analyze each stabilization mechanism separately.
\begin{figure}[h]
  \centering
  \includegraphics[width=478 px]{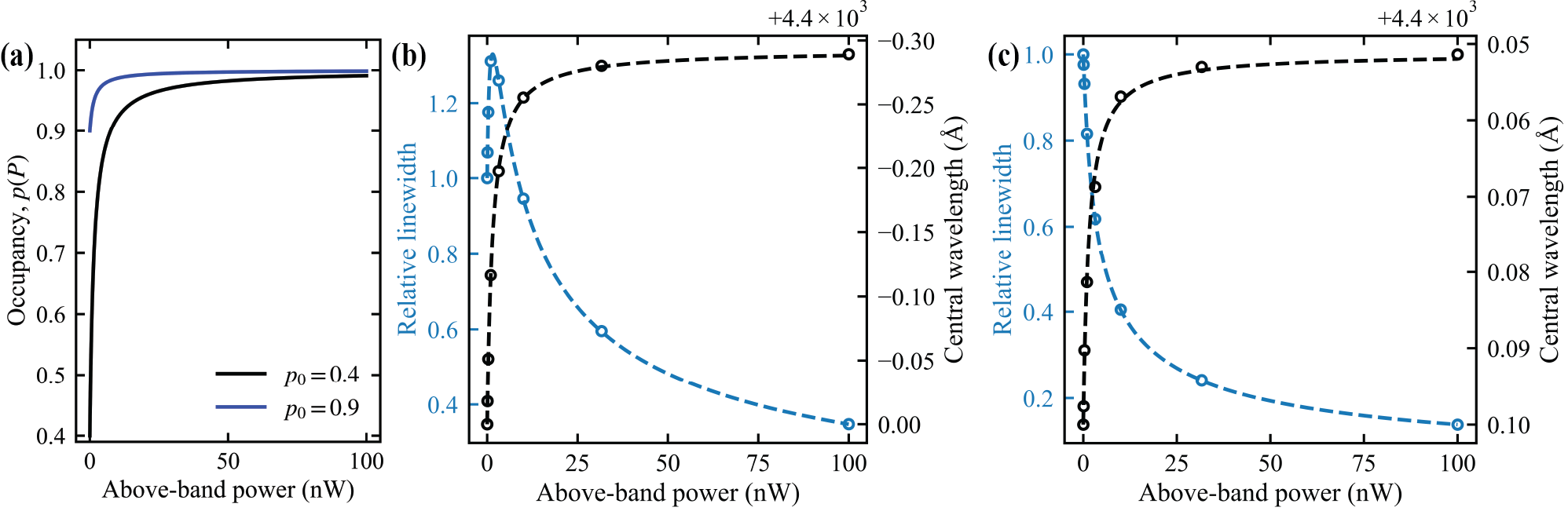}
  \caption{Optical suppression of trap noise with initial occupancy of (b) 0.4 and (c) 0.9}
  \label{fig:S5}
\end{figure}

\begin{figure}[h]
  \centering
  \includegraphics[width=480 px]{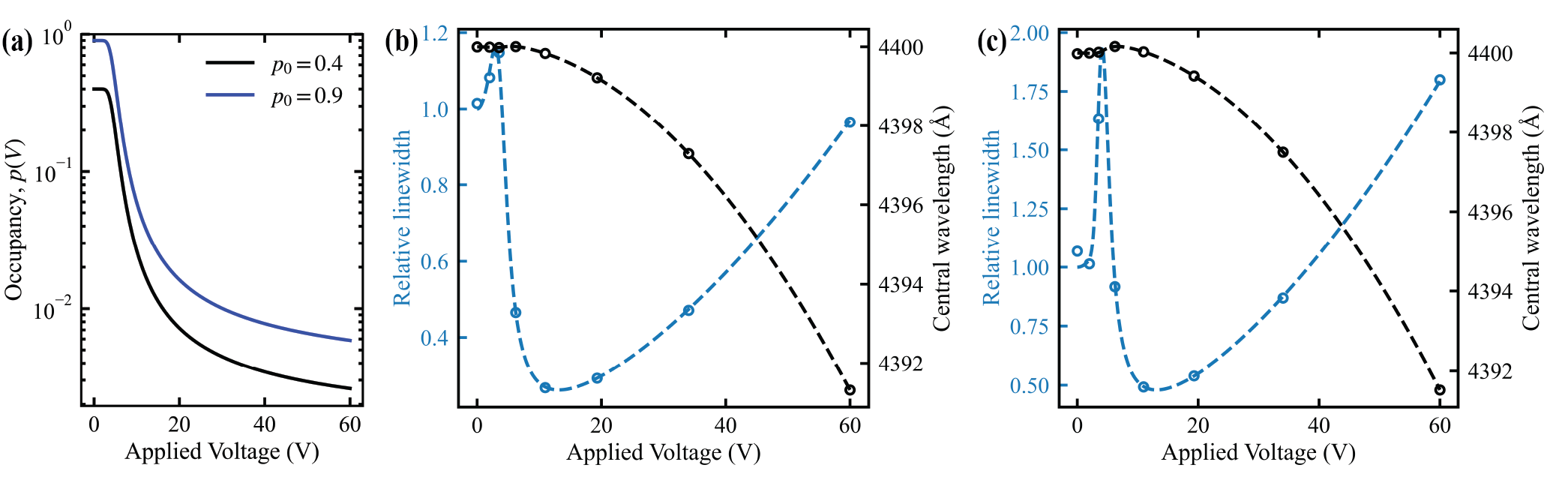}
  \caption{Electric suppression of trap noise with initial occupancy of (b) 0.4 and (c) 0.9}
  \label{fig:S6}
\end{figure}

\begin{itemize}
    \item For optical suppression of trap states, we compare the simulations with the analytical expressions in Fig.~\ref{fig:S5}. The dots are values extracted from the Monte Carlo simulation running under two scenarios. The calculation uses a saturation above-band power of \SI{1.5}{\nano\watt}, \(N=50\) traps within \SIrange{3}{8}{\nano\meter} of the donor, initial occupancies \(p_0=0.4\) and \(0.9\), and a final occupancy driven to \(1\). For the central wavelength shift we use a polarizability \( 1.44\times10^{-6}\ \mathrm{meV}\,(\mathrm{cm}/\mathrm{kV})^{2}\) and initial wavelength of \SI{440}{\nano m}. The dashed curves give the analytical result, Eqs.~\eqref{eq:sigmaP_final} and \eqref{eq:centerP_final}, in good agreement with the simulations. Above-band excitation raises the occupancy toward unity, and for initial \(p < 0.5\) the linewidth first broadens and then narrows as \(p \to 1\).
    \item For electric suppression of trap states, we compare the simulations with the analytical expressions in Fig.~\ref{fig:S6}. The dots are values extracted from the Monte Carlo. The calculation uses a trap depth of $E_*=800\  \mathrm{kV}/\mathrm{cm}$, relative field rate of $B = 50$ and $\alpha = 0.2$ and $\gamma = 1$. We use $3.3\times 10^6$ V/m as the conversion factor between voltage and field, based on section eq. \ref{eq:VtoFloc}. The number of traps and geometry is similar to the optical suppresion case with initial occupancies \(p_0=0.4\) and \(0.9\). The dashed curves give the analytical result, Eqs.~\eqref{eq:sigma_with_Erates} and \eqref{eq:mu_with_Erates}, in good agreement with the simulations. Field decreases the occupancy toward zero, and for initial \(p > 0.5\) the linewidth first broadens and then narrows as \(p \to 0\). For \(p < 0.5\), there is a slight broadening effect due to almost constant occupancy at the beginning. At higher fields for both cases, the quadratic coupling between field and noise dominates, showing a monotonic broadening behavior.
\end{itemize}

\subsection{Fit to experimental data}
In order to fit the closed form solution to the experimental data, we may estimate some of the parameters such that the variable space becomes smaller. This is important to enable the optimization program to find the local/global minimum. 

Based on $E_\ast=\tfrac{4\sqrt{2m^\ast}}{3\hbar q}\,\Phi^{3/2}$, we can estimate the characteristic field. For ZnSe, based on effective electron mass of $m^\ast\approx 0.16~ m_e$ \cite{10.1063/1.1713761} and an average trap depth of $\Phi \approx 0.5$ eV \cite{karczewski1994deep,hernandez1996investigation}, we can calculate $E_\ast\approx 972$ kV/cm.

For the distribution range and number of traps, we can assume the presence of traps starting immediately outside of Bohr radius $a^\ast = a_H \epsilon_r \frac{m_e}{m^\ast}$, which gives $a^\ast \approx 3$ nm. For the cutoff distance, at about $2a^\ast$, the contribution of field decays to about 6$\%$, possibly could be a valid approximation. To estimate the number of contributing traps, we use a wide range of effective trap densities $10^{12}-10^{14}\ \mathrm{cm}^{-2}$ \cite{seghier1997interface,sodhi2024surface,PhysRevB.89.125201}, considering the spectrum of single trap blinking or telegraph effect to ensemble trap wandering \cite{pelton2007evidence}. Based on this, we may estimate $N\approx 1-100$ traps. 
To fully address the broadening observed in Figure 2b in the main text, we add an additional $CE_0^2$ term for possible heating at high fields. This will ease the process of fit without enforcing the physical values to compensate for unrelated effects to trap states.

\subsection{Appendix: Isotropic Assumptions }
\begin{equation}
\langle u_{i} \rangle = 2 f_i E_b \Big\langle \cos(\phi_i) \Big\rangle
= \frac{2 f_i E_b}{2\pi} \int_0^{2\pi}\! \cos(\phi_i)\,d\phi_i = 0
\label{eq:angle_u}
\end{equation}
\begin{equation}
\langle u_{i}^2 \rangle = 4 f_i^2 E_b^2 \Big\langle \cos^2(\phi_i) \Big\rangle
= \frac{2 f_i E_b}{2\pi} \int_0^{2\pi}\! \cos^2(\phi_i)\,d\phi_i = 2 f_i^2 E_b^2
\label{eq:angle_u2}
\end{equation}
\begin{align}
\langle b_{ij} \rangle
&= 2 f_i f_j \Big\langle \cos(\phi_j-\phi_i) \Big\rangle
= \frac{2 f_i f_j}{(2\pi)^2} \int_0^{2\pi}\!\!\int_0^{2\pi}\! \cos(\phi_j-\phi_i)\,d\phi_i d\phi_j \nonumber\\
&= \frac{2 f_i f_j}{2\pi}\int_0^{2\pi}\! \cos u \, du = 0,
\label{eq:angle_b}
\end{align}
\begin{align}
\langle b_{ij}^2 \rangle
&= 4 f_i^2 f_j^2 \Big\langle \cos^2(\phi_j-\phi_i) \Big\rangle
= \frac{4 f_i^2 f_j^2}{(2\pi)^2} \int_0^{2\pi}\!\!\int_0^{2\pi}\! \cos^2(\phi_j-\phi_i)\,d\phi_i d\phi_j \nonumber\\
&= \frac{4 f_i^2 f_j^2}{2\pi} \int_0^{2\pi} \cos^2 u \, du
= 2 f_i^2 f_j^2,
\label{eq:angle_b2}
\end{align}
and for mixed products with a shared index,
\begin{align}
\langle b_{ij} b_{ik} \rangle
&= 4 f_i^2 f_j f_k \, \Big\langle \cos(\phi_j-\phi_i)\cos(\phi_k-\phi_i)\Big\rangle \nonumber\\
&= \frac{4 f_i^2 f_j f_k}{(2\pi)^3} \int_0^{2\pi}\!\!\int_0^{2\pi}\!\!\int_0^{2\pi}
\cos(\phi_j-\phi_i)\cos(\phi_k-\phi_i)\,d\phi_i d\phi_j d\phi_k \nonumber\\
&= \frac{4 f_i^2 f_j f_k}{2\pi} \int_0^{2\pi} \cos u \, du \cdot \frac{1}{2\pi}\int_0^{2\pi} \cos v \, dv = 0,
\label{eq:angle_mixed_zero}
\end{align}
since the two cosines are independent, zero-mean functions of their angles. and any mixed average with a single cosine vanishes, e.g.\ $\langle u_k b_{ij}\rangle=0$. 

\bibliographystyle{ieeetr}   
\bibliography{refsSI.bib}          
